\documentclass[draft]{ectaart}
\usepackage[dvips]{graphics}
\usepackage{epsfig}
\usepackage[psamsfonts]{amssymb}
\usepackage{amsmath}
\newcommand{\bs}{\boldsymbol}
\usepackage{natbib}
\usepackage{theorem}
\theoremheaderfont{\sc}
\newtheorem{theorem}{Theorem}[section]

{\theorembodyfont{\rm} \newtheorem{rem}{Remark}[section]}
{\theorembodyfont{\sf} \newtheorem{defs}{Definition}[section]}
{\theorembodyfont{\upshape} \newtheorem{exam}{Example}[section]}
\begin{document}

\title{Most Efficient Homogeneous Volatility Estimators}
\author{A. Saichev, D. Sornette, V. Filimonov}
\address{Department of Management, Technology and Economics,
ETH Zurich, Kreuzplatz 5, CH-8032 Zurich, Switzerland}

\maketitle

\begin{abstract}
We present a new theory of homogeneous volatility (and variance) estimators for
arbitrary stochastic processes. The main tool of our theory is the parsimonious encoding
of all the information contained in the OHLC prices for a given time interval by the
joint distributions of the high-minus-open, low-minus-open and close-minus-open values,
whose analytical expression is derived exactly for Wiener processes with drift.
The efficiency of the new proposed estimators is favorably compared with that of the
Garman-Klass, Roger-Satchell and maximum likelihood estimators.
\end{abstract}


\section{Introduction}

Volatility, defined as the standard deviation of the increments of the log-price over a specific time interval, is a universally used risk indicator. While the growing availability of high-frequency tick-by-tick price time series
has permitted the development of new efficient volatility estimators (see, for instance, Yang and Zhang (2000), Corsi et al. (2001), Andersen et al. (2003), A\"it-Sahalia (2005), Zhang et al. (2005)), most historical time series as well as databases of price time series, for the many tens of thousands of assets (stocks, commodities, bonds, currencies, derivatives and so on)
that exit worldwide, only record price in time steps coarse-grained for convenience (which is often daily). However, it is common practice that not just one (close) price is recorded for a given time step, but four of them, called the open-high-low-close (OHLC) of the price for that given interval. It is natural to exploit these four recorded values per time step to develop
better volatility estimators.

Rather than just using the time series of close-prices, here, we present
a comprehensive theory of homogeneous volatility (and variance) estimators of
arbitrary stochastic processes that fully exploit the OHLC prices.  For this, we develop
the theory of most efficient point-wise homogeneous OHLC volatility estimators, valid
for any price processes. We introduce the ``quasi-unbiased estimators'',
that can address any type of desirable constraints. The main tool of our theory is the parsimonious encoding
of all the information contained in the OHLC prices for a given time interval
in the form of general ``diagrams'' associated with the
joint distributions of the high-minus-open, low-minus-open and close-minus-open values. The diagrams
can be tailored to yield the most efficient estimators associated to any statistical properties of the underlying log-price stochastic process. Applied to Wiener processes for log-prices with drift, we provide explicit analytical expressions
for the most efficient point-wise volatility and variance estimators, based on the analytical expression of the
joint distribution of the high-minus-open, low-minus-open and close-minus-open values.

Our work improves on the following papers.
Garman and Klass (G\&K) (1980) introduced a quadratic estimator
for the variance of the Wiener process with drift for the log-price, which has rather low variance
but which is biased from non-zero drifts. Parkinson (1980) proposed a simple
quadratic variance estimator proportional to $(H-L)^2$, which is using only
a part of the information available from  OHLC prices.
Rogers and Satchell (R\&S) (1991,1994)
introduced another quadratic estimator for the variance of the Wiener process with drift, which is
unbiased for all drifts and has a larger fixed variance for all drifts equal to the variance of the process.
Both G\&K and R\&S estimators are focused on the variance, and do not present
estimators for the volatility, which is of obvious interest for financial applications.
Furthermore, the variance of their estimators is not provided for non-zero drifts.
Magdon-Ismail and Atiya (2003) obtain a maximum likelihood (ML) volatility estimator
based on the joint distribution of the high and low, previously obtained
by Domin\'e (1996). Their estimator does not use the close price
and is thus less efficient than the ML estimator using the full
information of the OHLC, as shown here. In addition, we will show that the ML estimator
is not the most efficient. Yang and Zhang (2000)
produced an unbiased and efficient quadratic
variance estimator, taking into account the OHLC of log-prices for $n>1$ consecutive days.
Their main novelty is to take into account the possible existence of jumps (or gaps) of prices
from yesterday's close till today's open prices. Their minimization of the
variance of their estimators requires the estimation of expectations of a quadratic
form of the OHLC which they only partly achieve due to the lack of knowledge
of the full joint distribution, which we offer in the Appendix.
Chan and Lien (2003) compared the empirical effectiveness of four estimators, the Parkinson, the G\&K and R\&S ones,
and the naive excursion range $H-L$ estimator.
In sum, the present paper can be viewed as providing the full underpinning theory
of all these previous works, since we are able to express efficient estimators
in the presence of arbitrary constraints from the explicit knowledge of the joint distribution
of the OHLC log-prices.

The paper is organized as follows. Section~2 describes the properties of the stochastic processes
for which our theory of most efficient homogeneous estimators is developed. Section~3 derives the general expressions for the most efficient volatility and variance OHLC estimators. Section~4 provides detailed analytical results on the statistical properties of the most efficient homogeneous estimators, for the case of Wiener process with drift. Section~5
compares the exact analytical results with those obtained using numerical simulations of millions
of realizations of the Wiener process with drift. Section~6 concludes. The Appendix
presents the joint probability density function of the
high-minus-open, low-minus-open and close-minus-open values
for the Wiener process with drift.

\section{Volatility of stochastic processes: models, definitions and properties}

\subsection{Volatility of simple stochastic log-price process}

The goal of this paper is to construct efficient estimators for the volatility of log-price processes.
First, we specify the general properties of the stochastic processes to which our estimators will be applied.

Let us consider the stochastic process $X(t)$, which is interpreted as the log-price of some asset at time $t$.  Its volatility over the time interval of duration $T_0$ is by definition the standard deviation of the increment $X(t_0+T_0)-X(t_0)$. We assume that $X(t)$ has stationary increments. Accordingly, for simplicity but without loss of generality, we can take $t_0=0$ and $X(0)=0$ and choose the time scale such that $T_0=1$.  All the rest of the paper is based on the
following definition of volatility:
\begin{defs} The volatility $\sigma$ of the stochastic process $X(t)$ is equal to the square-root of the variance of its increment per unit time
$$
\sigma= \sqrt{D}~ , \qquad  D = \text{Var}[X(1)]~ .
$$
\end{defs}

The estimators of the volatility $\sigma$ and of the variance $D\equiv\sigma^2$ will be denoted respectively $\hat{\sigma}$ and $\hat{D}$.

We consider the following class of  stochastic processes
\begin{equation}\label{genmod}
X(t) = \sigma A(t,\gamma)~,
\end{equation}
where $A(t,\gamma)$ is an auxiliary stochastic process, whose statistical properties are assumed to be known for any given value of the parameter $\gamma$. We assume additionally that the expectation and the variance of the stochastic process $A(t,\gamma)$ are finite:
$$
\text{E}[A(t,\gamma)]<\infty ~ , \qquad \sigma^2_0(t,\gamma) = \text{Var}[A(t,\gamma)]<\infty~ .
$$
It follows from \eqref{genmod} and from the definition of volatility that the stochastic process $A(t,\gamma)$
has a unity volatility: $\sigma_0(1,\gamma) = 1$.

Let us introduce the following auxiliary stochastic process
\begin{equation}\label{xtoy}
Y(t) = {X(t) \over \sigma_0(T,\gamma)} = \sigma B(t,\gamma) ~ , \qquad B(t,\gamma) = {A(t,\gamma) \over \sigma_0(T,\gamma)}~ .
\end{equation}
By construction, the variance of the increments of the ``normalized'' stochastic process $Y(t)$ over a time interval of arbitrary duration $T$ coincides with the variance of the increments of the original process $X(t)$ over the unit time interval of duration $T_0=1$:
$$
\text{Var}[B(T,\gamma)] = 1 \qquad \Rightarrow \qquad
\text{Var}[Y(T,\gamma)] = \sigma^2~ .
$$

Let us consider particular examples of the stochastic processes $X(t)$
given by \eqref{genmod} and of the corresponding $Y(t)$ defined by \eqref{xtoy}:

\begin{exam} The simplest and most common log-price process is the Wiener process
\begin{equation}\label{xwindrift}
X(t) = \mu t + \sigma W(t)~ ,
\end{equation}
where $W(t)$ is the standard Wiener process, such that
$\text{E}[W(t)]=0$, $\text{Var}[W(t)]= t$, while $\mu$ is the drift parameter. In this case, $\sigma_0(T,\gamma) = \sqrt{T}$, so that the auxiliary stochastic process $Y(t)$ \eqref{genmod} takes the form
$$
Y(t) = X(t) \big/ \sqrt{T} = \sigma B(t,\gamma)~ ,
$$
where
\begin{equation}\label{vtwiener}
B(t,\gamma) = v(\tau,\gamma) ~ , \qquad v(\tau,\gamma) = \gamma \tau + W(\tau) ~ .
\end{equation}
Here, we introduced the ``normalized'' time $\tau$ and the parameter $\gamma$:
\begin{equation}\label{gamdef}
\tau = {t\over T}~ , \qquad \gamma = {\mu \over \sigma} \sqrt{T}~ .
\end{equation}
\end{exam}

\begin{rem} In practical applications, the parameter $\gamma$, which is is proportional to the drift of the stochastic process $X(t)$ \eqref{xwindrift}, is generally unknown. Our strategy is to proceed in two steps: (i) determine the most
efficient volatility and variance estimators for a fixed value of $\gamma$, say $\gamma_0$; (ii) explore in details the efficiency of the estimators for values of  $\gamma$ that deviate from $\gamma_0$.
\end{rem}

\begin{exam} Let $X(t)$ be defined at discrete times $t=0,1,2,\dots$ and let it satisfy to recurrent relation
$$
X(n+1) = X(n)+ \mu + \sigma \epsilon_n~ , \quad X(0) = 0~ , \quad
n=0,1,2,\dots
$$
where $\{\epsilon_n\}$ is a sequence of iid random variables with zero expectation and unit variance $\text{Var}[\epsilon_n] = 1$. In order to estimate the volatility $\sigma$ from recorded values of $X(n)$ over a
discrete time interval of duration $N$, it is convenient to introduce the ``normalized'' discrete-time process
$$
Y(n) = {X(n) \over \sqrt{N}} = \sigma v(n,\gamma)~ ,
$$
where
\begin{equation}\label{vdiscret}
\begin{array}{c}
v(n,\gamma) = \gamma n + \omega(n)~ , \qquad n=1,2,\dots~ , \qquad v(0,\gamma) = 0~ ,
\\[2mm]\displaystyle
\gamma = {\mu \over  \sigma \sqrt{N}}~ , \qquad \omega(n) = {1 \over \sqrt{N}} \sum_{k=1}^n \epsilon_k~ .
\end{array}
\end{equation}
\end{exam}

\begin{rem}
If the random variables $\{\epsilon_k\}$ are Gaussian, the stochastic process $X(n)$ can be interpreted as the discrete-time version of the process $X(t)$ defined by \eqref{xwindrift}. More interesting is the case where $\{\epsilon_k\}$ are non-Gaussian random variables, with a fat tail probability density distribution $f(x)\sim |x|^{-1-p}$ for large $|x|$, with $p>2$ ensuring that the variance exists (see McKenzie, (2006) for an excellent review of the history of fat tails in financial returns).
\end{rem}

\subsection{OHLC volatility and variance estimators}

Given the observed realization of the stochastic process $X(t)$ within some time interval $t\in(0,T)$
over $m$ points, $(t_1,t_2,\dots,t_{m-1})\in (0,T), \quad t_m=T$, the most general expression of the estimator of
the volatility of $X(t)$ is the function
$$
\hat{\sigma}_m =\hat{\sigma}(X_1,X_2,\dots,X_m)~ ,
$$
of the recorded values
$$
X_1=X(t_1)~ , \quad X_2=X(t_2)~ , \quad \dots , \quad X_m = X(T)~ .
$$

Of particular interest for its widespread use and parsimonious representation
of a given realization of the process $X(t)$ over a finite time interval is the case $m=3$
that corresponds, in particular, to OHLC estimators. The four letters OHLC stand
respectively for Open, High, Low and Close. In the following, we focus on this case due to its
special significance, while it is understood that one can generalize the theory developed
here to higher-order multipoint estimators corresponding to any value $m>3$.

Without loss of generality, we pose $X(0)=0$ (in practice, the
relevant quantities are simply decreased by the opening value at time $=0$). Then, the high, low and close values of
a given realization of the stochastic process $X(t)$ within the time interval $t\in(0,T)$ are
\begin{equation}\label{hlcsup}
H = \sup_{t\in(0,T)} X(t)~ , \qquad L = \inf_{t\in(0,T)} X(t)~ , \qquad C=X(T)~ .
\end{equation}
\begin{defs} Among all three-points volatility and variance estimators, the \emph{OHLC estimators}
are defined as functions of only the three measures (high, low and close) of the realization of the
stochastic process $X(t)$ within the time interval $t\in(0,T)$ defined by (\ref{hlcsup}).
Specifically, OHLC volatility and variance estimators are functions which can be written as follows:
\begin{equation}\label{ohlcsanddest}
\hat{\sigma} =\hat{\sigma}(H,L,C)~ , \qquad \hat{D} =\hat{D}(H,L,C)~ .
\end{equation}

\end{defs}
Such OHLC estimators are well-known to be more efficient than the equidistant three-points estimators
corresponding to $t_k=k T/3$ ($k=1,2,3$).

\subsection{Quadratic OHLC variance and volatility estimators}

Almost all known variance OHLC estimators are quadratic forms of $H, L$ and $C$. Let us introduce
the vector  $\boldsymbol{X}^T=(H,L,C)$, where $^T$ denotes the transpose operation. Let us denote
by $\boldsymbol{Q}$ any positive-definite $3\times 3$ matrix.

\begin{defs} The variance OHLC estimator $\hat{D}$ is called \emph{quadratic} if it can be expressed as a quadratic form
\begin{equation}\label{quadvarest}
\hat{D} = {1 \over \sigma_0^2(T,\gamma)}~ \bs{X}^T \bs{Q} \bs{X} = \bs{Y}^T \bs{Q} \bs{Y} ~ , \qquad \text{where} \qquad \bs{Y} = {\bs{X} \over \sigma_0(T,\gamma)}~ .
\end{equation}
In turn, the volatility OHLC estimator $\hat{\sigma}$ is called  \emph{quadratic} if it can be represented in the form
\begin{equation}\label{quadvolest}
\hat{\sigma} = {1 \over \sigma_0(T,\gamma)}~\sqrt{\bs{X}^T \bs{Q} \bs{X}} = \sqrt{\bs{Y}^T \bs{Q} \bs{Y}} ~ .
\end{equation}
\end{defs}

Two well-known OHLC estimators are quadratic, as shown in the two examples \ref{ergjwak} and \ref{ynthnbad}.

\begin{exam} Rogers and Satchell  (1991) have suggested the following quadratic OHLC variance estimator
\begin{equation}\label{RS}
\hat{D}_\text{RS} = {1\over T}\, \left[H (H-C) + L (L-C) \right]~ .
\end{equation}
We will refer to this estimator as the \emph{R}\&\emph{S} \emph{variance estimator}. The corresponding expression\begin{equation}\label{RSvol}
\hat{\sigma}_\text{RS} = {1\over \sqrt{T}}\, \sqrt{H (H-C) + L (L-C)}
\end{equation}
will be called the \emph{R}\&\emph{S} \emph{volatility estimator}.
\label{ergjwak}
\end{exam}

The R\&S variance estimator \eqref{RS} has the nice property of being unbiased. Namely, for the Wiener process
defined by \eqref{xwindrift}, and for any $\sigma$ and $\mu$ (i.e. for any values of the parameter $\gamma$),
the expected value of the R\&S variance estimator \eqref{RS} is equal to the variance of the original process over the
time interval $[0,1]$: $\text{E}[\hat{D}_\text{RS}] = \sigma^2$.

\begin{exam} Another quadratic OHLC variance estimator  was suggested by Garman and Klass (1980). This \emph{G}\&\emph{K} \emph{variance estimator} is defined by
\begin{equation}\label{GKvar}
\hat{D}_\text{GK} = {1 \over T} \left[k_1\, (H-L)^2 - k_2\, (C(H+L)- 2 H
L) -k_3\, C^2\right]~ ,
\end{equation}
where
$k_1 = 0.511$, $k_2 = 0.019$, $k_3 = 0.383$.
The square root of expression \eqref{GKvar} will be referred to as the  \emph{G}\&\emph{K} \emph{volatility estimator}.
\label{ynthnbad}
\end{exam}
For the Wiener process \eqref{xwindrift}, the G\&K variance estimator is unbiased only
if the drift is equal to zero. In general,
$\text{E}[\hat{D}_\text{GK}] \neq \sigma^2$ if $\mu\neq 0$ ($\gamma\neq 0$). This bias is a shortcoming of the G\&K variance estimator. Its advantage is that, for zero drift $\mu=0$ ($\gamma=0$), its variance is significantly
smaller than the variance of the R\&S variance estimator.

\subsection{Homogenous variance and volatility estimators}

In order to more clearly understand the key properties of any quadratic estimators,
it is instructive to introduce ``generalized'' R\&S and G\&K estimators for the
general stochastic process $X(t)$ defined by \eqref{genmod}.
For definiteness, we will focus here on the ``generalized''R\&S variance estimator
obtained by replacing $T$ by $\sigma_0^2(T,\gamma)$ in \eqref{RS}:
$$
\hat{D}_\text{RS} = {1\over \sigma_0^2(T,\gamma)}\, \left[H (H-C) + L (L-C) \right]~ .
$$
Using relations \eqref{xtoy}, it can be written in the form
\begin{equation}\label{Dtod}
\hat{D}_\text{RS} = \sigma^2 \hat{d}_\text{RS}~ , \qquad
\hat{d}_{\text{RS}} = \bar{H} (\bar{H}-\bar{C}) + \bar{L} (\bar{L}-\bar{C})~ ,
\end{equation}
where $\hat{d}_{\text{RS}}$ is function of the high, low and close values of the auxiliary stochastic process $B(t,\gamma)$ defined by \eqref{xtoy} within the interval $t\in(0,T)$:
\begin{equation}\label{hlcbar}
\bar{H} = \sup_{t\in(0,T)} B(t,\gamma)~ , \qquad \bar{L} = \inf_{t\in(0,T)} B(t,\gamma)~ , \qquad \bar{C}=B(T,\gamma)~ .
\end{equation}
Accordingly, the R\&S volatility estimator is equal to
\begin{equation}\label{RSvolhom}
\hat{\sigma}_\text{RS} = \sigma \hat{s}_\text{RS}~ , \qquad \hat{s}_\text{RS} = \sqrt{\bar{H} (\bar{H}-\bar{C}) + \bar{L} (\bar{L}-\bar{C})}~ .
\end{equation}

The R\&S variance estimator $\hat{D}_\text{RS}$ given by \eqref{Dtod} has the following important property: It is equal to the product of the (unknown) variance $\sigma^2$ of the original process $X(t)$ defined by \eqref{genmod} and of the random factor $\hat{d}_\text{RS}$. The statistical properties of $\hat{d}_\text{RS}$ are expressed via the statistical properties of auxiliary process $B(t,\gamma)$, which are known by definition.  Therefore, for a given $\gamma$, the statistical properties of $\hat{d}_\text{RS}$ do not depend on the variance $\sigma^2$ of the original process $X(t)$ defined in \eqref{genmod}. Moreover, since the R\&S variance estimator is unbiased, the expectation of $\hat{d}_\text{RS}$ is equal to unity: $\text{E}[\hat{d}_{\text{RS}}] \equiv 1$.
Correspondingly, one can quantitatively characterize the relative error of the R\&S variance estimator by the variance of the factor $\hat{d}_{\text{RS}}$,
$$
\text{Var}[\hat{d}_{\text{RS}}] = \text{E}[\hat{d}^2_{\text{RS}}]-1~ ,
$$
which does not depend (for a given $\gamma$) on the sought variance $\sigma^2$. Figuratively speaking, one can interpret relation \eqref{Dtod} as if the sought variance $\sigma^2$ was known while its random factor $\hat{d}_\text{RS}$
was unknown. Thus, the R\&S variance estimator is all the more efficient, the smaller is the variance of factor $\hat{d}_\text{RS}$.

\begin{defs} If the OHLC variance estimator is represented in the form
\begin{equation}\label{genhomosc}
\hat{D} = \sigma^2 \hat{d}
\end{equation}
where the factor
\begin{equation}\label{canvarest}
\hat{d} = \hat{d}(\bar{H},\bar{L}, \bar{C})
\end{equation}
depends only on the high, low and close values \eqref{hlcbar} of the auxiliary stochastic process $B(t,\gamma)$ and does not depend (for any given $\gamma$) on the variance $\sigma^2$ of the original stochastic process $X(t)$, then we refer to the factor $\hat{d}$ \eqref{canvarest} as the \emph{canonical variance estimator}. Similarly, if the volatility OHLC estimator is represented in the following form, analogous to \eqref{RSvolhom},
\begin{equation}\label{escanvol}
\hat{\sigma} = \sigma \hat{s}~ , \qquad \hat{s}=\hat{s}(\bar{H},\bar{L}, \bar{C})~ ,
\end{equation}
then the factor $\hat{s}$ is the \emph{canonical volatility estimator}.
\label{tyhyhnuj5i7k5i7m}
\end{defs}

Obviously, the estimators \eqref{genhomosc} and \eqref{escanvol} are \emph{unbiased}, for a given $\gamma=\gamma_0$, if
\begin{equation}\label{expdteqone}
\text{E}[\hat{d}|\gamma_0]=1~ , \qquad \text{E}[\hat{s}|\gamma_0]=1~ .
\end{equation}
Here and below, we use the notations $\text{E}[\dots|\gamma_0]$, $\text{Var}[\dots|\gamma_0]$
for conditional expectations and variances, under the condition that the parameter $\gamma$ is equal to $\gamma_0$.

\begin{rem}
In general, all volatility $\hat{\sigma}$ and variance $\hat{D}$ estimators \eqref{ohlcsanddest} are functions of $H$, $L$ and $C$. However, it is not true that all of them accept canonical estimators $\hat{s}$ and $\hat{d}$ \eqref{canvarest}, \eqref{escanvol}, depending on the variables $\bar{H}$, $\bar{L}$, $\bar{C}$ \eqref{hlcbar}. In the present paper, we explore only \emph{homogeneous} estimators, defined below, which are expressed via canonical estimators.
\end{rem}

\begin{defs}

The OHLC variance estimator is called homogeneous if it can be represented in the form
\begin{equation}\label{volesthomfone}
\hat{D}(H,L, C) =   h_2 (H,L,C)\big/ \sigma_0^2(T,\gamma)~ ,
\end{equation}
where $h_2 (H,L,C)$ is a second-order homogeneous function. Analogously, the volatility estimator is called homogeneous if it can be represented in the form
\begin{equation}\label{homogvoldefone}
\hat{\sigma}(H,L,C) =  h_1 (H,L,C)\big/ \sigma_0(T,\gamma)~ ,
\end{equation}
where $h_1$ is a first-order homogeneous function.
\end{defs}

\begin{theorem}\label{th1}
The homogeneous OHLC variance estimators $\hat{D}(H,L, C)$ \eqref{volesthomfone} accept
the representation form \eqref{genhomosc}, \eqref{canvarest}.
\end{theorem}

\emph{\textbf{Proof.}}
It follows from \eqref{hlcsup}, \eqref{hlcbar} and definition \eqref{xtoy} of the auxiliary stochastic process $B(t,\gamma)$, that the following equalities are true
\begin{equation}
H = \alpha\bar{H}~ , \quad L = \alpha \bar{L}~ , \quad C = \alpha \bar{C}~ , \quad \alpha =  \sigma \sigma_0(T,\gamma)~.
\label{kiol6kiteh}
\end{equation}
Thus, one can rewrite relation \eqref{volesthomfone} in the form
$$
\hat{D}(H,L, C) =
h_2 (\alpha \bar{H},\alpha \bar{L},\alpha \bar{C})\big/ \sigma_0^2(T,\gamma)~ .
$$
From the homogeneity property of the second order homogeneous function $h_2$,
$$
h_2(\alpha \bar{H}, \alpha \bar{L}, \alpha \bar{C}) \equiv \alpha^2 h_2(\bar{H}, \bar{L}, \bar{C})~ ,
$$
we obtain
$$
\hat{D}(H,L,C) = \sigma^2 h_2(\bar{H}, \bar{L}, \bar{C})~ .
$$
Thus, the homogeneous estimators \eqref{volesthomfone} are reduced to \eqref{genhomosc} and \eqref{canvarest}, where $\hat{d}= h_2(\bar{H},\bar{L}, \bar{C})$.
$\square$

\begin{rem} One can prove in a similar way that homogenous volatility estimators \eqref{homogvoldefone} are reduced to the form \eqref{escanvol}.
\end{rem}

\begin{defs} The variance \eqref{volesthomfone} and volatility \eqref{homogvoldefone} estimators are the \emph{most efficient} \emph{homogeneous estimators}, for a given $\gamma_0$,  if the corresponding canonical variance and volatility estimators satisfy relations \eqref{expdteqone}, while their variances
$$
\text{Var}[\hat{d}|\gamma_0] = \text{E}[\hat{d}^2|\gamma_0]-1~ , \quad \text{Var}[\hat{s}|\gamma_0] = \text{E}[\hat{s}^2|\gamma_0]-1 ~ ,
$$
are the smallest among the variances of all possible canonical homogeneous estimators, which are unbiased at $\gamma_0$,
\end{defs}

\begin{rem} All quadratic estimators are homogeneous. This results from their definition \eqref{quadvarest},
since the quadratic form $\bs{X}^T \bs{Q} \bs{X}$ is a second order homogeneous function of its argument $\bs{X}$. Analogously, the quadratic volatility estimator \eqref{quadvolest} is homogeneous, because $\sqrt{\bs{X}^T \bs{Q} \bs{X}}$ is a homogeneous function of order one. In particular, the quadratic R\&S \eqref{RS} and G\&K \eqref{GKvar} variance estimators are homogeneous.
\end{rem}

More insight in homogeneous OHLC estimators can be obtained by
representing $(H,L,C)$ in the following ``spherical'' (or geographic) coordinates
which embody parsimoniously the homogeneity property:
\begin{equation}\label{geogrback}
\begin{array}{c}
H = R \cos\Theta \cos\Phi~ , \quad L = R \cos\Theta \sin\Phi~ , \quad C = R \sin\Theta~ ,
\\
R= \sqrt{H^2 +L^2 + C^2}~ ,
\\ \displaystyle
\Theta = \arctan\left({C \over \sqrt{H^2 + L^2}}\right) ~, \quad \Phi = \arctan\left({L \over H} \right)~.
\end{array}
\end{equation}

\begin{theorem} Any variance estimator of the form
\begin{equation}\label{varspherest}
\hat{D}= {R^2 \over \sigma_0^2(T,\gamma)} \varphi(\Theta,\Phi)~ ,
\end{equation}
where $R, \Theta$ and $\Phi$ are given by (\ref{geogrback}) and
$\varphi(\theta,\phi)$ is an arbitrary function, is a homogeneous variance estimator.
Reciprocally, any homogenous variance estimator \eqref{volesthomfone} can be expressed in the form (\ref{varspherest}).
Similarly,
\begin{equation}\label{volspherest}
\hat{\sigma}= {R \over \sigma_0(T,\gamma)} \psi(\Theta,\Phi)~ ,
\end{equation}
where $\psi(\theta,\phi)$ is arbitrary function, is a homogeneous volatility estimator and reciprocally.
\end{theorem}

\emph{\textbf{Proof.}} It follows from \eqref{geogrback} that $R^2$ is a second order homogeneous function of its arguments $(H,L,C)$, while $\Theta$ and $\Phi$ are zero order homogeneous functions of the same arguments. Accordingly, $\varphi(\Theta,\Phi)$ is a zero order homogenous function of $(H,L,C)$, while $R^2 \varphi(\Theta,\Phi)$ is a second order homogeneous function of $(H,L,C)$. Thus, due to theorem~\ref{th1}, the estimator \eqref{varspherest} is represented in homogeneous form as
\begin{equation}\label{canbobtbtrvarest}
\begin{array}{c}
\hat{D} = \sigma^2 \hat{d}(\bar{H}, \bar{L}, \bar{C})~ , \qquad \text{where} \qquad \hat{d}(\bar{H}, \bar{L}, \bar{C}) = \bar{R}^2 \varphi(\Theta, \Phi)~ ,
\\[1mm] \displaystyle
\bar{R} = \sqrt{\bar{H}^2+ \bar{L}^2+ \bar{C}^2} ~ .
\end{array}
\end{equation}
In turn, it is obvious that any homogeneous estimator \eqref{volesthomfone}, is represented in the form \eqref{varspherest}, where
$$
\varphi(\Theta, \Phi) = h_2(\cos\Theta \cos\Phi, \cos\Theta \sin\Phi, \sin\Theta)~ .
$$

Using a similar derivation, it is easy to prove that $\hat{\sigma}$ given by \eqref{volspherest} is a homogeneous volatility estimator, i.e.,
\begin{equation}\label{varcanestthph}
\hat{\sigma}= \sigma \hat{s}((\bar{H}, \bar{L}, \bar{C}) ~ , \quad \text{where} \quad \hat{s}= \bar{R} \psi(\Theta, \Phi)~ . \qquad \square
\end{equation}

\begin{rem}
The inequalities
$$
\bar{L} \leqslant \bar{C} \leqslant \bar{H}~ , \qquad \bar{H}\geqslant 0~ , \qquad \bar{L}\leqslant 0~ ,
$$
resulting from the definition of $H, L, C$, impose
that $\bar{R}$, $\Theta$ and $\Phi$ should satisfy
$$
\begin{array}{c}\displaystyle
0\leqslant\bar{R} <\infty~ , \quad -{\pi \over 2} \leqslant \Phi \leqslant 0~ , \quad s(\Phi) \leqslant \Theta \leqslant c(\Phi) ~ , \\[2mm]
s(\phi) = \arctan(\sin\phi)~ , \qquad c(\phi) = \arctan(\cos\phi)~ .
\end{array}
$$
\end{rem}

\begin{defs} We will refer to the functions $\varphi(\theta,\phi)$ and $\psi(\theta,\phi)$, defined respectively by \eqref{canbobtbtrvarest} and \eqref{varcanestthph}, as the \emph{diagrams} of the homogeneous OHLC variance and volatility estimators.
\label{yjujifkd}
\end{defs}

\begin{exam} From the definitions \eqref{RS} and \eqref{GKvar}, the diagrams of the R\&S and G\&K variance estimators are
\begin{equation}\label{rsdiagr}
\begin{array}{c}\displaystyle
\varphi_\text{RS}(\theta,\phi) = \cos^2 \theta - {1 \over 2} \sin 2 \theta ( \cos\phi+ \sin \phi)~,
\\[1mm]\displaystyle
\varphi_\text{GK}(\theta,\phi) = k_1 \cos^2\theta (\cos\phi - \sin \phi)^2 \\[1mm] \displaystyle
+ k_2 \left[ \cos^2 \theta \sin 2\phi - {1 \over 2} \sin 2 \theta (\cos \phi + \sin \phi) \right] - k_3 \sin^2 \theta~ .
\end{array}
\end{equation}
\end{exam}

It is probable that R\&S and G\&K estimators are not the most efficient quadratic estimators for any given value $\gamma_0$ of the parameter $\gamma$. It is therefore natural to search for the most efficient  quadratic estimators, at a given value $\gamma_0$, which might be more efficient than R\&S and G\&K estimators. We will determine below the most efficient homogeneous volatility and variance OHLC estimators for any given $\gamma_0$. The following theorem summarizes the relations between the most efficient quadratic and the most efficient homogeneous OHLC estimators.

\begin{theorem} Let $\text{Var}_q[\hat{d}|\gamma_0]$ and $\text{Var}_q[\hat{s}|\gamma_0]$ be the variances of the most efficient quadratic canonical OHLC estimators for a given $\gamma_0$. Let $\text{Var}_h[\hat{d}|\gamma_0]$ $\text{Var}_h[\hat{s}|\gamma_0]$ be the variances of the most efficient homogeneous canonical OHLC estimators for the same given $\gamma_0$. Then, the following inequalities hold true
\begin{equation}\label{qhvaresineqs}
\text{Var}_q[\hat{d}|\gamma_0] \geqslant \text{Var}_h[\hat{d}|\gamma_0] ~ , \qquad \text{Var}_q[\hat{s}|\gamma_0] \geqslant \text{Var}_h[\hat{s}|\gamma_0] ~ .
\end{equation}
In another words, at a given value $\gamma_0$, the most efficient homogeneous OHLC estimator is no less efficient than the most efficient quadratic OHLC estimator.
\end{theorem}

\emph{\textbf{Proof.}} Denoting as $\Omega_q$ the set of quadratic OHLC estimators, and as $\Omega_h$ the set of homogeneous OHLC estimators, we have $\Omega_q \subset \Omega_h$. The inequalities  \eqref{qhvaresineqs}
derive from this inclusion. $\square$

\section{Diagrams of most efficient OHLC homogeneous estimators}

In this section, we derive the expressions for the most efficient (at $\gamma=\gamma_0$) homogeneous variance and volatility OHLC estimators , whose canonical estimators are given by expressions \eqref{canbobtbtrvarest} and \eqref{varcanestthph}. To make clear that these estimators depend on $\gamma_0$, we will use the following notations for the diagrams of the most efficient homogeneous estimators: $\varphi(\theta,\phi; \gamma_0)$ and  $\psi(\theta,\phi; \gamma_0)$.

We assume the existence of the joint probability density function (pdf) $\bar{\mathcal{Q}}(h,l,c;\gamma)$
of the random variables $(\bar{H},\bar{L},\bar{C})$ given by equalities \eqref{hlcbar}. The pdf
$\bar{\mathcal{Q}}(h,l,c;\gamma)$ depends on the parameter $\gamma$.
The pdf $\bar{\mathcal{Q}}(h,l,c;\gamma)$ is defined by
\begin{equation}\label{pdfhlcdef}
\begin{array}{c}
\bar{\mathcal{Q}}(h,l,c;\gamma)d h dl dc =
\\[2mm]
\Pr\{\bar{H}\in (h,h+dh), \bar{L}\in (l,l+dl), \bar{C}\in (c,c+dc)\}~ ,
\end{array}
\end{equation}
which expresses the probability that  $(\bar{H},\bar{L},\bar{C})$ take specific values to within
infinitesimal intervals. The Appendix gives the explicit expression of
the pdf $\bar{\mathcal{Q}}(h,l,c;\gamma)$ for the special case
of the Wiener process $v(\tau,\gamma)$ defined in \eqref{vtwiener}.

Let us consider first the canonical variance estimator
\begin{equation}\label{varcanohlcgamzest}
\hat{d} = \bar{R}^2 \varphi(\Theta, \Phi;\gamma_0)~ .
\end{equation}
The diagram of this estimator can be written as
\begin{equation}\label{mevardiagnorm}
\varphi(\theta, \phi;\gamma_0) = { G(\theta, \phi;\gamma_0) \over \text{E}[\mathstrut\bar{R}^2 G(\Theta, \Phi;\gamma_0)|\gamma_0]}~ ,
\end{equation}
where the function $G(\theta, \phi;\gamma_0)$ will be defined below. The expectation term in the denominator of expression\eqref{mevardiagnorm} is equal to
$$
\text{E}[\bar{R}^2 G(\Theta, \Phi;\gamma_0)|\gamma_0] = \int_{-\pi/2}^0 d\phi \int_{s(\phi)}^{c(\phi)} \cos\theta d\theta G(\theta,\phi;\gamma_0) g_2(\theta,\phi;\gamma_0)~ ,
$$
where
\begin{equation}\label{gnintrdef}
g_n(\theta,\phi;\gamma) = \int_0^\infty \rho^{2+n} \bar{\mathcal{Q}}(\rho \cos\theta\cos\phi, \rho \cos\theta \sin\phi,\rho \sin\theta;\gamma) d\rho~ .
\end{equation}

We stress the important property that the canonical OHLC variance estimator given by \eqref{varcanohlcgamzest} with \eqref{mevardiagnorm} is unbiased at $\gamma=\gamma_0$, since its expectation is
$$
\text{E}[\hat{d}|\gamma_0] = { \text{E}[\bar{R}^2 G(\Theta, \Phi;\gamma_0)|\gamma_0] \over \text{E}[\bar{R}^2 G(\Theta, \Phi;\gamma_0)|\gamma_0]} = 1~ .
$$
Thus, we look for the function $G(\Theta, \Phi;\gamma_0)$ that makes the unbiased canonical variance estimator \eqref{varcanohlcgamzest} with \eqref{mevardiagnorm} the most efficient for a given $\gamma_0$.

\begin{theorem} The diagram of the unbiased most efficient  homogeneous canonical variance estimator for a given $\gamma_0$ is equal to
\begin{equation}\label{mevardiagrest}
\varphi(\theta, \phi;\gamma_0) = {1 \over \mathcal{E}(\gamma_0)} ~ {g_2(\theta,\phi;\gamma_0) \over g_4(\theta,\phi;\gamma_0)}~ ,
\end{equation}
where $g_n(\theta,\phi;\gamma)$ is defined by expression (\ref{gnintrdef}) and
\begin{equation}\label{mathegamdef}
\mathcal{E}(\gamma) = \int_{-\pi/2}^0 d\phi \int_{s(\phi)}^{c(\phi)} \cos\theta d\theta~ {g_2^2(\theta,\phi;\gamma) \over g_4(\theta,\phi;\gamma)}~ .
\end{equation}
\end{theorem}

\emph{\textbf{Proof.}}
The variance of the unbiased homogeneous canonical estimator \eqref{varcanohlcgamzest} with \eqref{mevardiagnorm} is equal to
\begin{equation}\label{23}
\text{Var}\left[\hat{d}|\gamma_0\right] =
{\displaystyle\int_{-\pi/2}^0 d\phi
\int_{s(\phi)}^{c(\phi)} \cos\theta d\theta ~ G^2(\theta,\phi;\gamma_0)
g_4(\theta,\phi;\gamma_0)  \over \displaystyle\left(\int_{-\pi/2}^0
d\phi \int_{s(\phi)}^{c(\phi)} \cos\theta d\theta ~ G(\theta,\phi;\gamma_0)
g_2(\theta,\phi;\gamma_0) \right)^2} -1 ~ .
\end{equation}

We use the Schwarz inequality to determine the minimal value of the variance given by \eqref{23} of the canonical estimator. Omitting for the sake of conciseness the limits in the integrals, we represent the Schwarz inequality in the form
$$
\left(\iint A(\theta,\phi) B(\theta,\phi)d\theta d\phi\right)^2
\leqslant \iint A^2(\theta,\phi) d\theta d\phi \iint
B^2(\theta,\phi) d\theta d\phi~ ,
$$
where $A(\theta,\phi)$ and $B(\theta,\phi)$ are arbitrary real-valued functions. Taking here
$$
\begin{array}{c} \displaystyle
A(\theta,\phi) = G(\theta,\phi;\gamma_0)
\sqrt{g_4(\theta,\phi;\gamma_0) \cos \theta}~ ,
\\[4mm] \displaystyle
B(\theta,\phi) =
g_2(\theta,\phi;\gamma_0) \sqrt{{\cos \theta \over
g_4(\theta,\phi;\gamma_0)}}~ ,
\end{array}
$$
we obtain
$$
 \begin{array}{c} \displaystyle
   \left(\iint  G(\theta,\phi;\gamma_0) g_2(\theta,\phi;\gamma) \cos\theta d\theta d\phi \right)^2 \leqslant \\[4mm] \displaystyle
   \iint G^2(\theta,\phi;\gamma_0) g_4(\theta,\phi;\gamma_0) \cos\theta
   d\theta d\phi \iint {g_2^2(\theta,\phi;\gamma_0) \over
   g_4(\theta,\phi;\gamma_0)} \cos\theta d\theta d\phi~ .
 \end{array}
$$
It follows from \eqref{23} and from the last inequality that the variance of any canonical variance estimator satisfies the inequality
\begin{equation}\label{24}
\text{Var}\left[\hat{d}(\Theta, \Phi;\gamma_0)|\gamma_0\right] \geqslant V(\gamma_0)~, \qquad
V(\gamma) = {1 \over \mathcal{E}(\gamma)} -1~ ,
\end{equation}
where $\mathcal{E}(\gamma)$ is defined by expression (\ref{mathegamdef}).
Taking into account \eqref{mathegamdef}, \eqref{23}  and \eqref{24}, the variance of the canonical variance estimator reaches its minimal value $V(\gamma_0)$ for the following choice of the function $G(\theta,\phi;\gamma_0)$:
$$
G(\theta,\phi;\gamma_0) = {g_2(\theta,\phi;\gamma_0) \over g_4(\theta,\phi;\gamma_0)}~ .
$$
This corresponds to the diagram $\varphi(\theta,\phi;\gamma_0)$ given by expression \eqref{mevardiagrest}. $\square$

An analogous derivation provides the unbiased most efficient canonical volati\-lity estimator, for a given $\gamma_0$. The main corresponding results are summarized in the following theorem.

\begin{theorem} The diagram $\psi(\theta,\phi;\gamma_0)$ of the unbiased most efficient  homogeneous canonical OHLC volatility estimator,
defined by
\begin{equation}\label{canvolhomohlcgamz}
\hat{s} = \bar{R} \psi(\Theta,\Phi;\gamma_0)~,
\end{equation}
is equal to
\begin{equation}\label{mevardiagrestv}
\begin{array}{c}\displaystyle
\psi(\theta, \phi;\gamma_0) = {1 \over \mathcal{F}(\gamma_0)} ~ {g_1(\theta,\phi;\gamma_0) \over g_2(\theta,\phi;\gamma_0)}~ ,
\\[4mm]\displaystyle
\mathcal{F}(\gamma) = \int_{-\pi/2}^0 d\phi \int_{s(\phi)}^{c(\phi)} \cos\theta d\theta~ {g_1^2(\theta,\phi;\gamma) \over g_2(\theta,\phi;\gamma)}~ .
\end{array}
\end{equation}
The variance of the most efficient canonical OHLC volatility estimator is equal to
\begin{equation}\label{wgamleastvoldef}
W(\gamma_0) = {1 \over \mathcal{F}(\gamma_0)} -1~ .
\end{equation}
\end{theorem}

\begin{defs} $V(\gamma)$ defined in \eqref{24} is called the \emph{lowest bound} of the variance of the canonical variance estimator, for a given value of the parameter $\gamma$. Analogously, $W(\gamma)$ given by \eqref{wgamleastvoldef}
is called the \emph{lowest bound} of the variance of the canonical volatility estimator, for the given value of the parameter $\gamma$.
\end{defs}

\section{Properties of most efficient OHLC variance estimators for the Wiener process}

The Appendix derives the explicit expression of the pdf $\bar{\mathcal{Q}}(h,l,c;\gamma)$ of the high, low and close values of the Wiener process $v(\tau,\gamma)$ defined in \eqref{vtwiener}. This section uses this explicit knowledge to
explore the quantitative properties of the most efficient canonical estimators for this particular case and
compare them with those of the R\&S and G\&K canonical variance estimators.

\begin{quote}
\centerline{
\includegraphics[width=9cm]{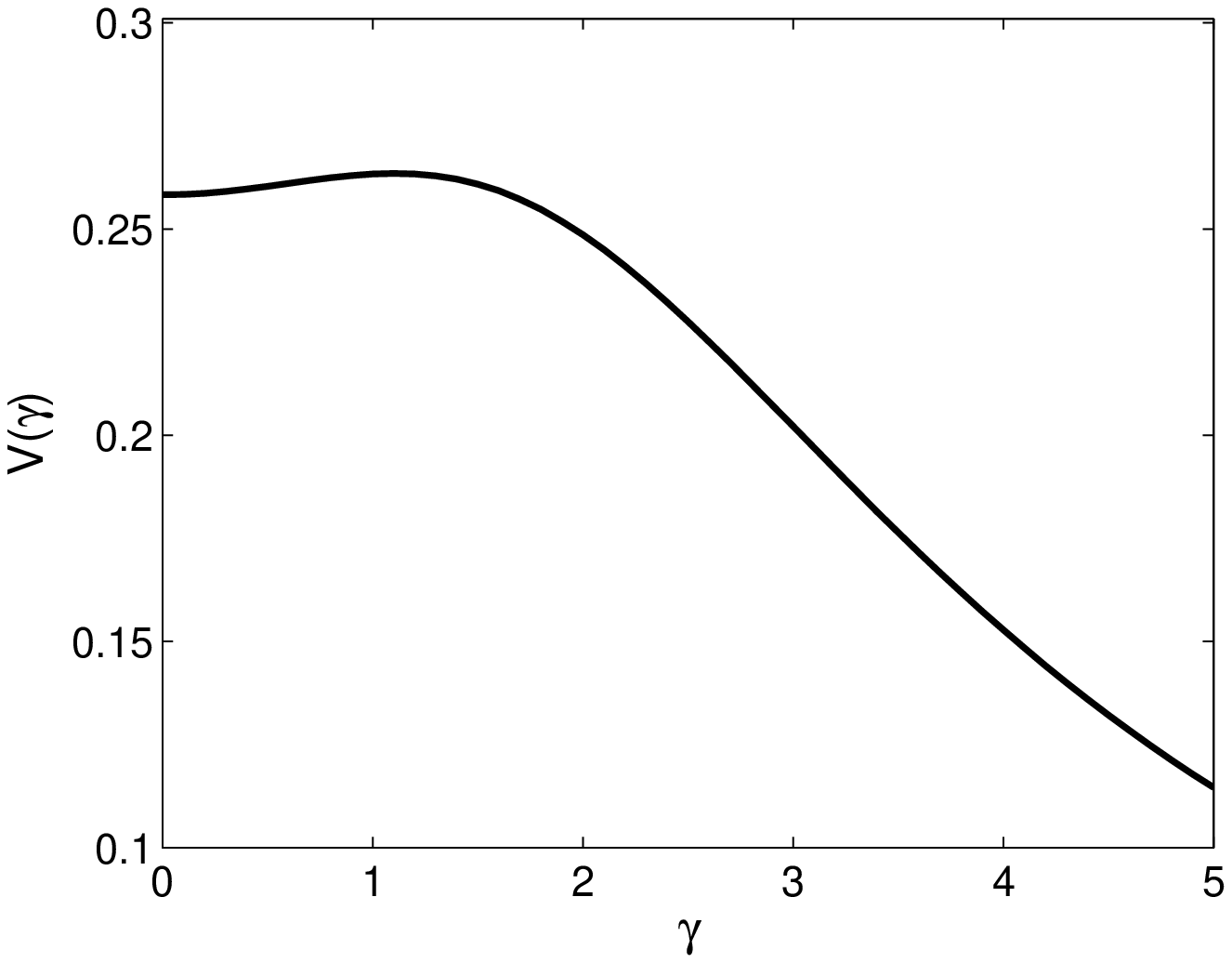}}
{\bf Fig.~1:} \small{Dependence of the lowest bound $V(\gamma)$ given by (\ref{24}) of the variance of the homogeneous canonical variance estimator as a function of $\gamma$. $V(0)=0.2583$.}
\end{quote}

\subsection{Variance of  canonical variance estimators}

Let us first consider the lowest bound $V(\gamma)$ given by \eqref{24} of the variance of the homogeneous canonical variance estimator. For the Wiener process model, it is easy to calculate numerically the function $V(\gamma)$, which
is represented in figure~1. The variance of the most efficient canonical variance estimator at $\gamma_0=0$ is equal to $V(0)\approx 0.258$, which can be compared with the corresponding variances for the G\&K and R\&S canonical variance estimators: $\text{Var}[\hat{d}_\text{GK}|0]\approx 0.27$, $\text{Var}[\hat{d_\text{RS}}|0]\approx 0.331$ (Rogers and Satchell, 1991). Thus, at $\gamma=0$, the G\&K variance estimator has almost the same efficiency as the most efficient (for $\gamma_0=0$) homogeneous variance estimator, while the efficiency of the R\&S estimator is significantly worse.
These results are reflecting the closeness of the diagrams of the G\&K and most efficient estimators, while the diagram of the R\&S estimator drastically differs from the diagram of most efficient estimator, as shown in figure~2.

\begin{quote}
\centerline{
\includegraphics[width=6.5cm]{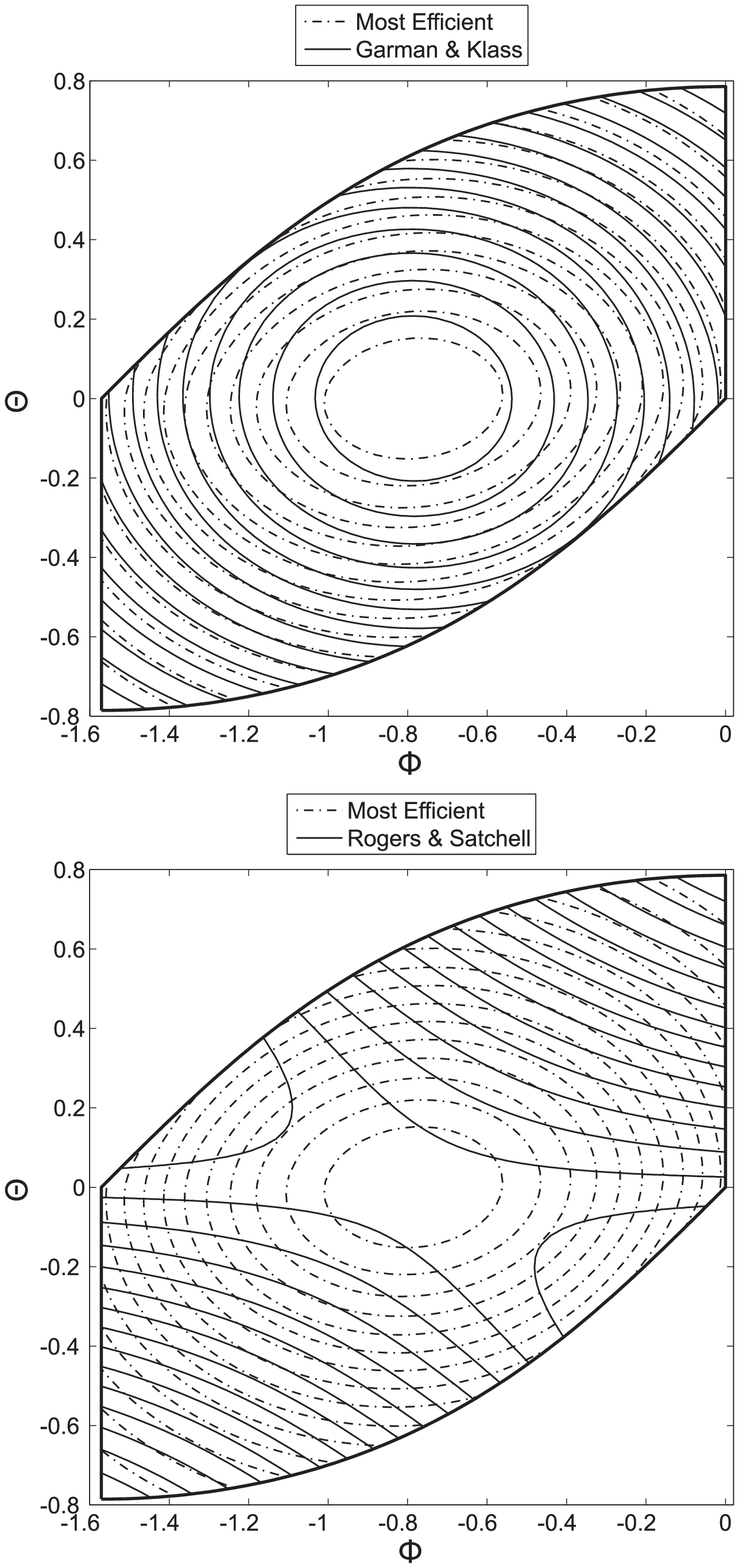}}
{\bf Fig.~2:} \small{Diagrams of the R\&S, G\&K and most efficient (for $\gamma_0=0$) variance estimators.
See definition \ref{yjujifkd} for the meaning and construction of the diagrams.}
\end{quote}

\subsection{Bias and efficiency of the most efficient ($\gamma_0$) variance estimator}

The homogeneous variance estimator with diagram \eqref{mevardiagrest} is unbiased and most efficient only for a given
value $\gamma_0$. In general, the value of $\gamma$ is unknown. It is thus necessary to
quantify the bias and efficiency of the homogeneous estimator for different values $\gamma \neq \gamma_0$, and compare it with the biases and efficiencies of the G\&K and R\&S variance estimators.

For this, we first determine the expected value and the variance of an arbitrary canonical
variance estimator given by \eqref{varcanohlcgamzest}.
Calculations similar to those performed in the previous section yield
\begin{equation}\label{26}
\begin{array}{c}
\text{E}\left[\hat{d}(\Theta,\Phi)|\gamma\right] =
\mathcal{K}_1(\gamma)~ , \quad \text{Var}\left[\hat{d}(\Theta,\Phi)|\gamma\right] =
\mathcal{K}_2(\gamma) - \mathcal{K}_1^2(\gamma)~ ,
\\[2mm]\displaystyle
\mathcal{K}_n(\gamma) = \displaystyle\int_{-\pi/2}^0 d\phi
\int_{s(\phi)}^{c(\phi)} \cos\theta d\theta~
g_{2n}(\theta,\phi;\gamma) \varphi^n(\theta,\phi)~.
\end{array}
\end{equation}
Substituting the expression \eqref{mevardiagrest} for the diagram of the most efficient estimator into
equation \eqref{26} yields
$$
\begin{array}{c}\displaystyle
\text{E}\left[\hat{d}(\Theta, \Phi;\gamma_0)|\gamma\right]=
{\mathcal{E}(\gamma,\gamma_0) \over \mathcal{E}(\gamma_0)}~ ,
\\[4mm] \displaystyle
\mathcal{E}(\gamma,\gamma_0) = \int_{-\pi/2}^0
d\phi \int_{s(\phi)}^{c(\phi)} \cos\theta d\theta~
{g_2(\theta,\phi;\gamma) g_2(\theta,\phi;\gamma_0)\over
g_4(\theta,\phi;\gamma_0)}~ .
\end{array}
$$

Figure~3 presents the dependence as a function of
$\gamma$ of the expected value of the most efficient canonical variance estimators given by \eqref{varcanohlcgamzest} with \eqref{mevardiagrest}. The expectations of the R\&S and G\&K canonical variance estimators, whose diagrams are given by \eqref{rsdiagr}, are also shown for comparison. While the R\&S variance estimator is unbiased for all $\gamma$'s, the most efficient estimators at $\gamma_0$ are unbiased only in the neighborhood of $\gamma=0$ and of $\gamma=\gamma_0$. Comparing the G\&K and the most efficient estimators, the homogeneous estimator, which is the most efficient for $\gamma_0=1$ for instance, is not significantly biased over the whole range
$0\leqslant\gamma\lesssim 1.5$ and remains much less biased than
the G\&K estimator over the range $0\leqslant\gamma\leqslant2$.

\begin{quote}
\centerline{
\includegraphics[width=9cm]{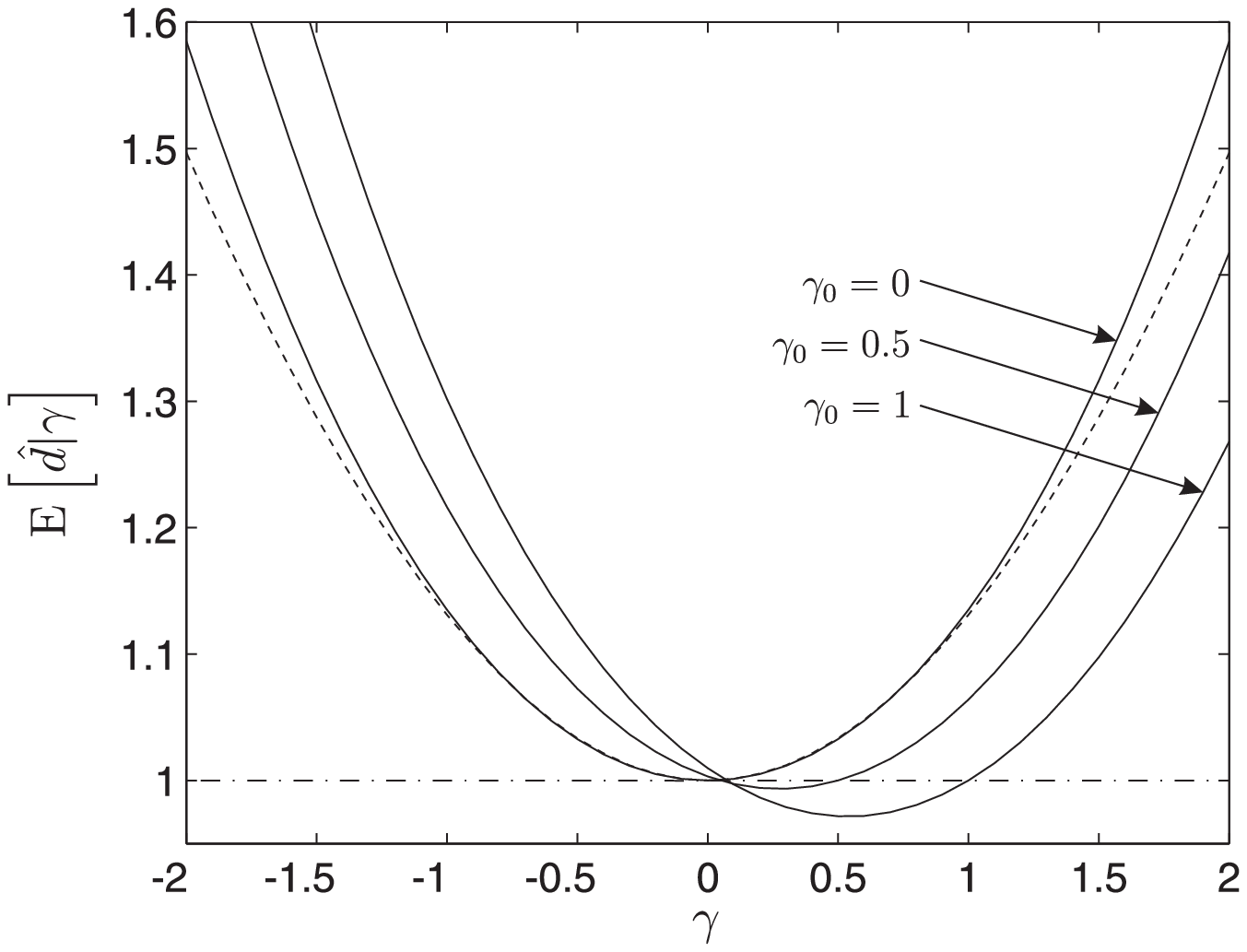}}
{\bf Fig.~3:} \small{Dependence as a function of $\gamma$ of the expected values of
the R\&S (dash-dot line) and G\&K (dashed line)
canonical variance estimators and of the most efficient variance estimators
for $\gamma_0=0;0.5;1$ (solid lines, top-down)}
\end{quote}

\begin{quote}
\centerline{
\includegraphics[width=9cm]{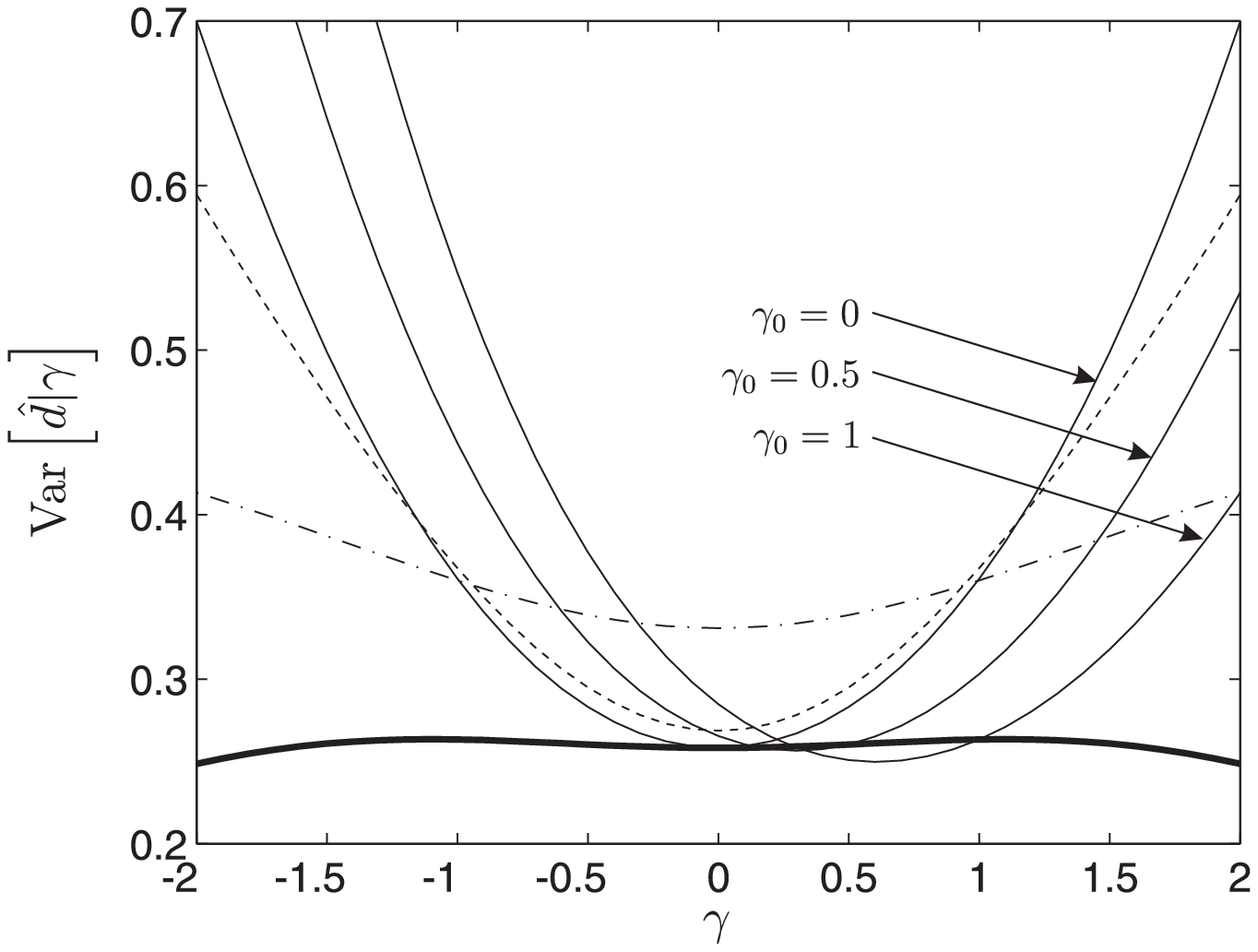}}
{\bf Fig.~4:} \small{Dependence as a function of $\gamma$ of the variances of
the R\&S (dash-dot line), G\&K (dashed line) and most efficient variance estimators for $\gamma_0=0;0.5;1$ (solid lines). The heavy solid line is the lowest bound variance given by \eqref{24}.}
\end{quote}

\begin{quote}
\centerline{
\includegraphics[width=9cm]{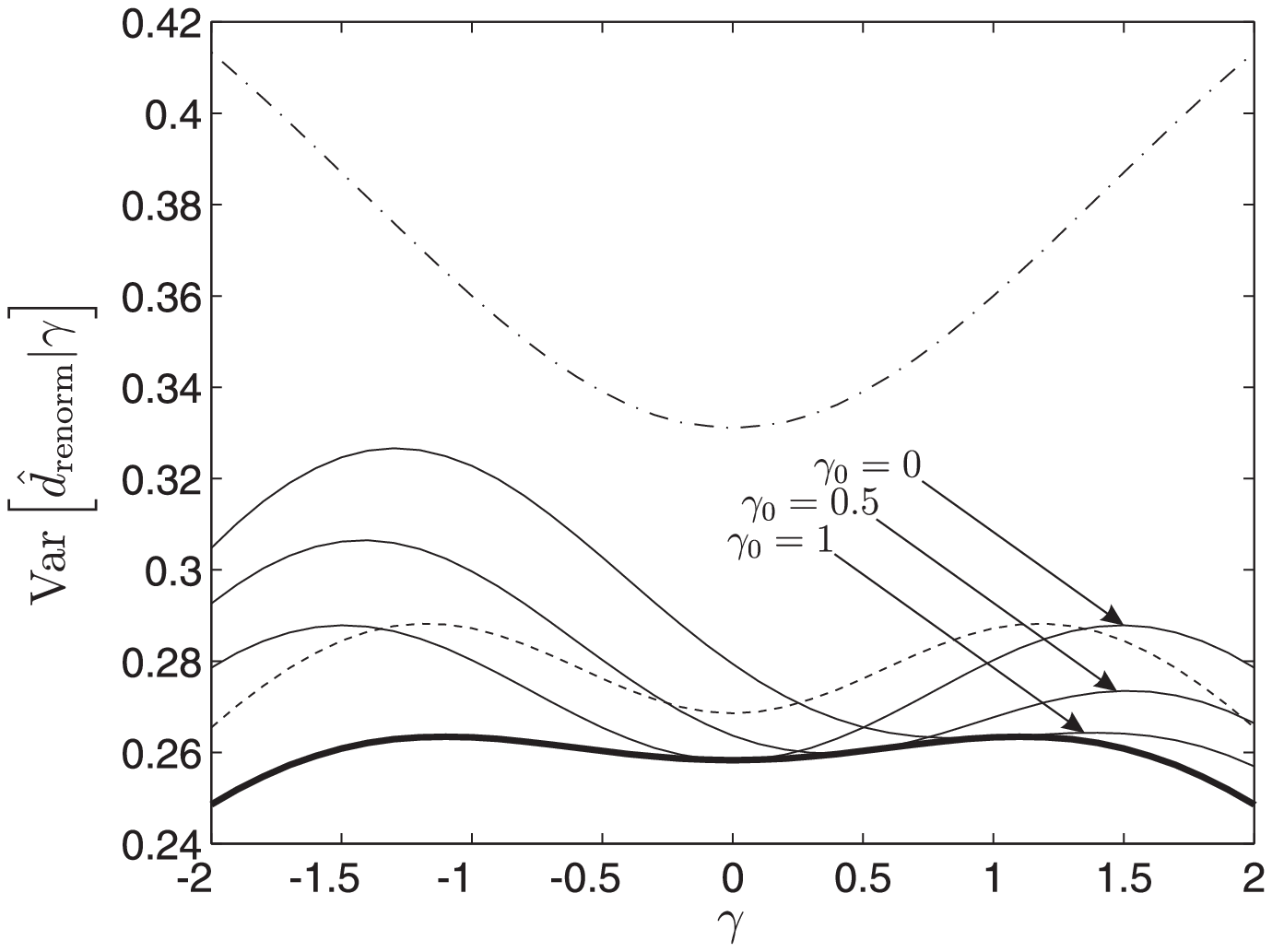}}
{\bf Fig.~5:} \small{Dependence as a function of $\gamma$ of the variances of
the \emph{renormalized} R\&S (dash-dot line) and G\&K (dashed line)
canonical variance estimators, and the most efficient estimators (solid lines), as defined
by expression (\ref{tbybkw;s}).}
\end{quote}

Calculation of the variance (for any $\gamma$) of
the canonical variance estimator, which is most efficient at $\gamma_0$, gives
$$
\begin{array}{c} \displaystyle
\text{Var}\left[\hat{d}(\Theta,
\Phi;\gamma_0)|\gamma\right] = {\mathcal{M}(\gamma,\gamma_0) -
\mathcal{E}^2(\gamma,\gamma_0) \over \mathcal{E}^2(\gamma_0) }~ ,
\\[4mm] \displaystyle
\mathcal{M}(\gamma,\gamma_0) = \int_{-\pi/2}^0 d\Phi
\int_{s(\phi)}^{c(\phi)} \cos\theta d\theta\,
{g_4(\theta,\phi;\gamma) g_2^2(\theta,\phi;\gamma_0)\over
g_4^2(\theta,\phi;\gamma_0)}~ .
\end{array}
$$

Figure~4 shows the dependence as a function of $\gamma$ of the variances of the R\&S and
G\&K canonical variance estimators and of the most efficient homogeneous variance
estimators for different $\gamma_0$. One can observe that the homogeneous variance estimator, which is the most efficient at $\gamma_0=1$, is both less biased and significantly more efficient than the G\&K estimator over the interval $0\lesssim \gamma \lesssim 2$.

One should not be surprised to observe in figure~4 several intervals along the $\gamma$ axis in which the variances of the estimators are smaller than the lower bound $V(\gamma)$ given by \eqref{24}. Indeed, the lower bound for the variance given by \eqref{24} is suitable only for unbiased estimators. Therefore, the  ``strange'' behavior of the variance plots does not mean that these estimators are more efficient than the most efficient estimator at the given $\gamma$, but rather that
they are biased at this point. With the proper renormalization
\begin{equation}
\hat{d}_\text{renorm}=\hat{d} \big/\text{E}\left[\hat{d}|\gamma\right]~,
\label{tbybkw;s}
\end{equation}
one can see that, for any $\gamma$ values, the estimators have variances
which are indeed bounded from below by the lower bound $V(\gamma)$, as shown in figure~5.

\subsection{Probabilistic properties of homogeneous estimators}

Knowing the exact explicit expression of the pdf $\bar{\mathcal{Q}}(h,l,c;\gamma)$ of the high, low and close values of the Wiener process $v(\tau,\gamma)$ defined in \eqref{vtwiener} given in the Appendix, we can
go beyond the calculations of the expectations and variances of the estimators described
in previous subsections and derive their full distribution.  In particular, the knowledge of the full
distribution of the estimators allows one to determine the confidence intervals of
the quasi-unbiased estimators introduced in section \ref{yhtnwss}.

Let us consider the pdf of the canonical variance estimator \eqref{canbobtbtrvarest}. For a given $\gamma$,
it is defined by the following expression
$$
f(u;\gamma) =
\text{E}\left[\delta(u-\bar{R}^2 \varphi(\Theta,\Phi))|\gamma\right]~ .
\label{yhnsjkvwflobkv}
$$
Using the standard properties of the delta-function of a composite argument, we can rewrite
the previous definition (\ref{yhnsjkvwflobkv}) in the form
$$
f(u;\gamma)= \text{E}\left[{1 \over \sqrt{u \varphi(\Theta,\Phi)}}\delta\left(\bar{R}- \sqrt{{u \over \varphi(\Theta,\Phi)}}\right)\big|\gamma \right]~ ,
$$
or more explicitly
\begin{equation}\label{varpdf}
\begin{array}{c} \displaystyle
f(u;\gamma) =
\frac{\sqrt{u}}2
\int_{-\pi/2}^0 d\phi \int_{s(\phi)}^{c(\phi)}
{\cos\theta d\theta \over \varphi^{3/2}(\theta,\phi)}
\times
\\[4mm] \displaystyle
\bar{\mathcal{Q}}\left(
\sqrt{\frac u{\varphi(\theta,\phi)}}\cos\theta\cos\phi,
\sqrt{\frac u{\varphi(\theta,\phi)}}\cos\theta\sin\phi, \sqrt{\frac u{\varphi(\theta,\phi)}}\sin\theta; \gamma
\right)~ .
\end{array}
\end{equation}

We use expression (\ref{varpdf}) to obtain, by numerical integration, the pdf's of
R\&S, G\&K and of the most efficient ($\gamma_0=0$) canonical variance estimators,
calculated for $\gamma=0$. These three pdf's are represented in figure~6.
\begin{quote}
\centerline{
\includegraphics[width=9cm]{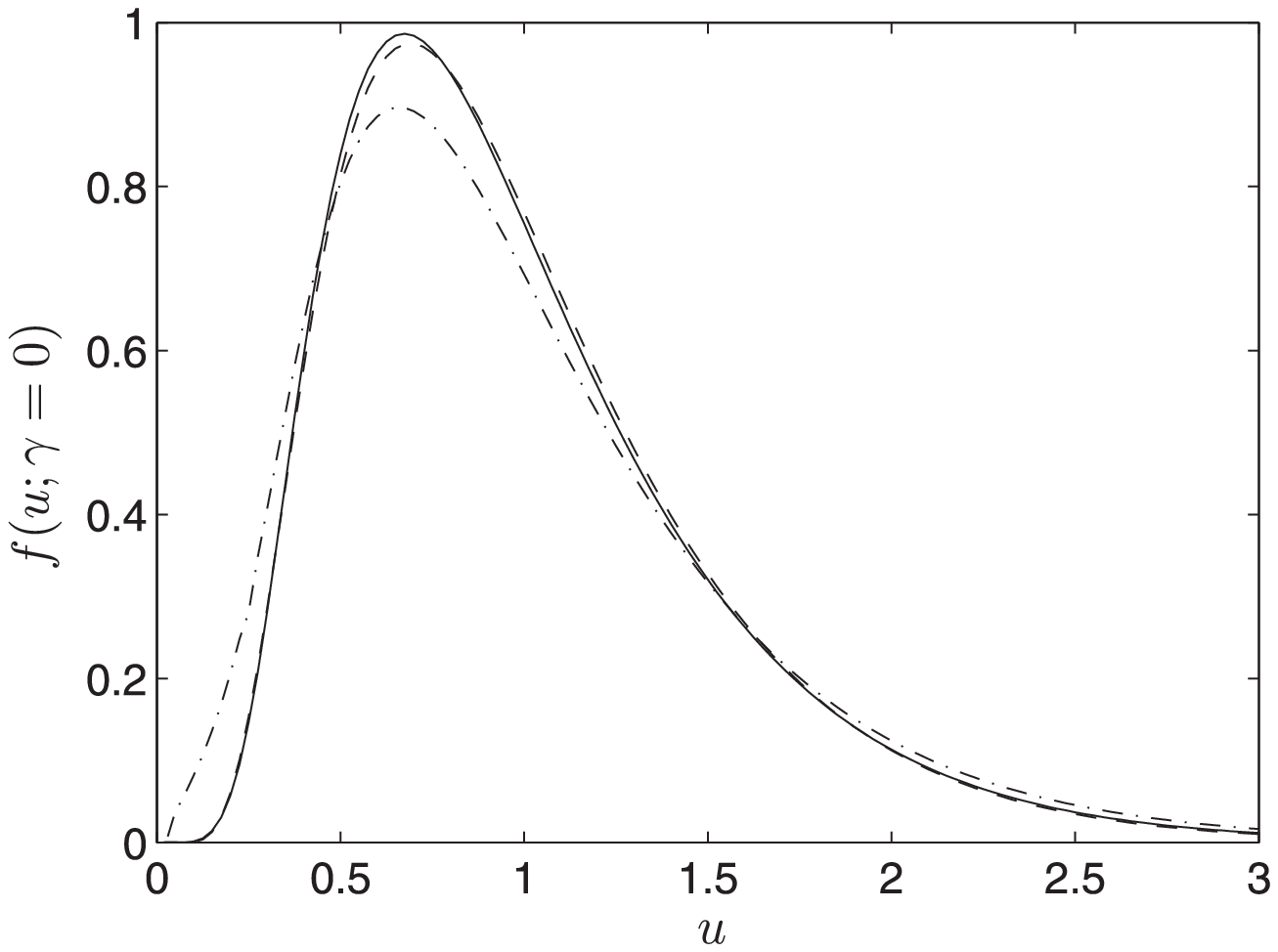}}
{\bf Fig.~6:} \small{Pdfs of the R\&S (dash-dot line),
  G\&K (dashed line) and of the most efficient ($\gamma_0=0$)
  canonical variance estimators (solid line), at $\gamma=0$.}
\end{quote}

\begin{quote}
\centerline{
\includegraphics[width=9cm]{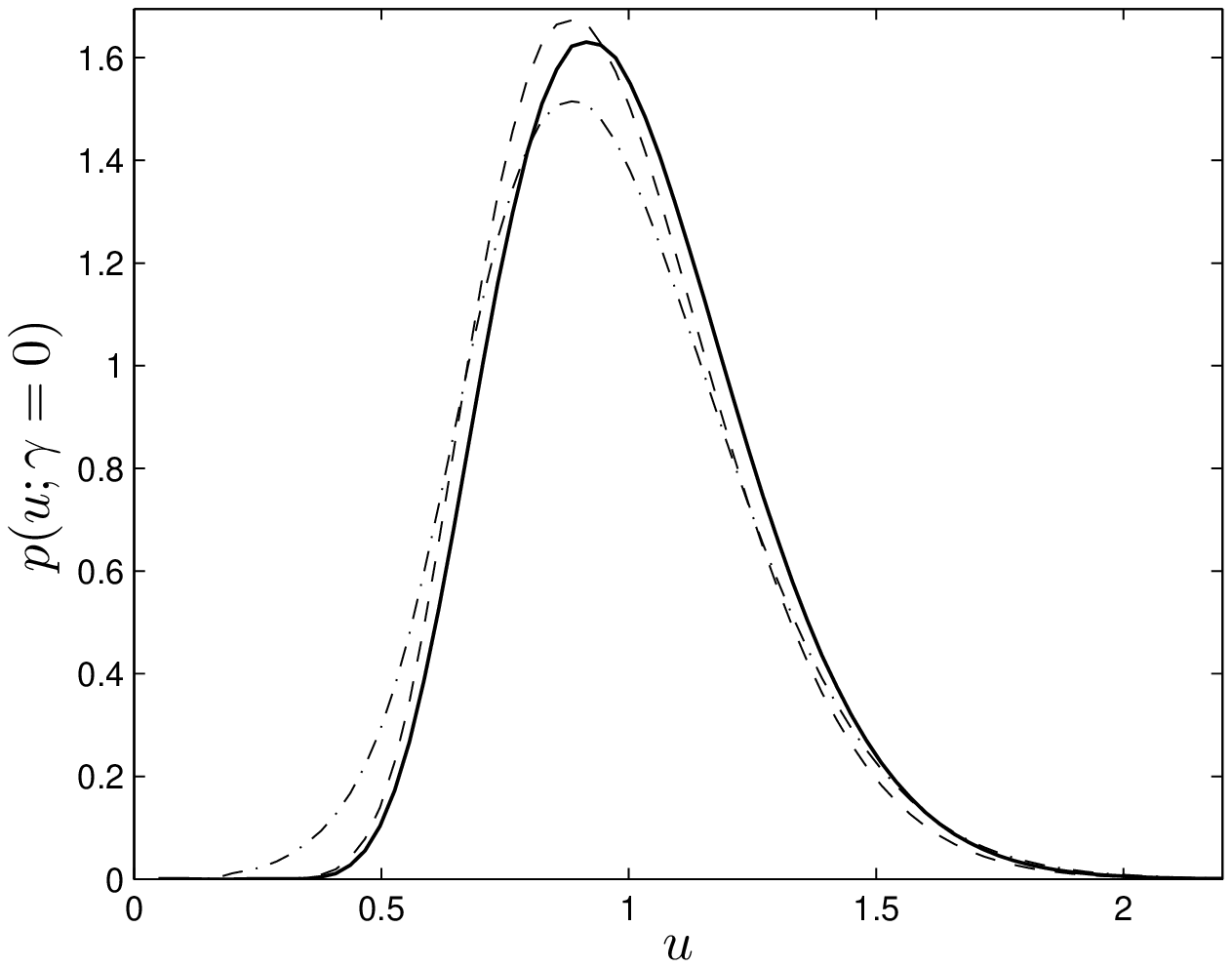}}
{\bf Fig.~7:} \small{Pdfs of the R\&S (dash-dot line),
  G\&K (dashed line) and of the most efficient ($\gamma_0=0$)
  canonical volatility estimators (solid line) at $\gamma=0$.}
\end{quote}

Similarly, the pdf of the canonical volatility estimator is defined by
$$
p(u;\gamma)= \text{E}[\delta(u-\bar{R} \psi(\Theta, \Phi))|\gamma]~,
$$
and its explicit expression, analogous to \eqref{varpdf} formula, reads
\begin{equation}\label{pugampdf}
\begin{array}{c} \displaystyle
p(u;\gamma) =
\\[4mm] \displaystyle
u^2
\int_{-\pi/2}^0 d\phi \int_{s(\phi)}^{c(\phi)}
{\cos\theta d\theta \over \psi^{3}(\theta,\phi)}
\bar{\mathcal{Q}}\left(
\frac {u \cos\theta\cos\phi}{\psi(\theta,\phi)},
\frac {u \cos\theta\sin\phi}{\psi(\theta,\phi)}, \frac {u \sin\theta}{\psi(\theta,\phi)}; \gamma
\right)~ .
\end{array}
\end{equation}
Figure~7 shows the pdf's given by \eqref{pugampdf} of the
R\&S, G\&K and of the most efficient ($\gamma_0=0$) canonical volatility estimators, for $\gamma=0$.

\subsection{Quasi-unbiased quasi-optimal estimators \label{yhtnwss}}

The previous subsections have made it clear that the most efficient unbiased ($\gamma_0$) estimators are not the most efficient for $\gamma\neq\gamma_0$, nor are they unbiased. Since varying $\gamma_0$ corresponds to scanning these most efficient estimators, which remain efficient in a neighborhood of their $\gamma_0$, this suggests to introduce
new reasonably efficient and approximately unbiased estimators, obtained as linear superpositions of the most efficient canonical homogeneous variance estimators:
\begin{equation}\label{39}
\hat{d}(\Theta,\Phi) = \bar{R}^2
\int_{-\infty}^\infty {h(\gamma_0) \over \mathcal{E}(\gamma_0)}\,
{g_2(\Theta,\Phi;\gamma_0) \over g_4(\Theta,\Phi;\gamma_0)}
d\gamma_0~ .
\end{equation}
Here, $h(\gamma_0)$ is some weighting function, whose explicit expression must be determined
from some optimization criterion. A possible requirement is that $h(\gamma_0)$ be such
as to both minimize the bias of the estimator \eqref{39} and maximize  its efficiency within some given $\gamma$ interval,
according to some criterion.

To demonstrate the principle of this approach, we search for the function $h_0(\gamma_0)$ that ensures that the estimator \eqref{39} is unbiased. The corresponding condition is that the expected value of the composed estimator \eqref{39} given by
$$
\text{E}\left[\hat{d}(\Theta,\Phi)|\gamma\right] =
\int_{-\infty}^\infty  {h(\gamma_0) \mathcal{E}(\gamma, \gamma_0)
\over \mathcal{E}(\gamma_0)} d\gamma_0~ ,
$$
be equal to $1$. Condition $\text{E}\left[\hat{d}(\Theta,\Phi)|\gamma\right] =1$
then provides an integral equation for the function $h_0(\gamma_0)$.
In practice, it is more convenient to look for \emph{quasi-unbiased} estimators,
which are exactly unbiased at $2K+1$ values of the parameter $\gamma$, for instance at
\begin{equation}\label{42}
\gamma_i = i\, {\Gamma \over K} ~ , \qquad i = -K,
-K+1, \dots -1, 0 , 1 , \dots, K-1, K ~ .
\end{equation}
Given these $2K+2$ constraints, it is natural to search for a solution
constructed as the sum of $2K+1$ most efficient ($\gamma_i$)
canonical variance estimators:
\begin{equation}\label{43}
\hat{d}(\Theta,\Phi) = \bar{R}^2 \sum_{i=-K}^K h_i
\varphi_i(\Theta,\Phi)~ , \quad
\varphi_i(\theta,\phi) = {1
\over \mathcal{E}(\gamma_i)}\, {g_2(\theta,\phi;\gamma_i) \over
g_4(\theta,\phi;\gamma_i)}~ .
\end{equation}
The $2K+1$ unknown coefficients $\{h_i, i=-K, ..., +K\}$ are to be determined
from the $2K+2$ constraints of an absence of bias at the discrete $\gamma$ values
(\ref{42}). We refer to $\Gamma$ as the \emph{band width} of the
quasi-unbiased estimator \eqref{43}, while $K$ is its \emph{order}.

In particular, the quasi-unbiased estimator of zero order corresponds
to the previously studied most efficient ($\gamma_0=0$) canonical variance
estimator. The first order quasi-unbiased estimator is equal to
\begin{equation}\label{45}
\hat{d}(\Theta,\Phi) = \bar{R}^2
\left[h_{-1}\varphi_{-1}(\Theta,\Phi)+
h_{0}\varphi_{0}(\Theta,\Phi)+h_{1}\varphi_{1}(\Theta,\Phi) \right] ~ ,
\end{equation}
and so on.

The expected value of the quasi-unbiased estimator \eqref{43} is equal to
\begin{equation}
\text{E}\left[\hat{d}(\Theta,\Phi)|\gamma\right] = \sum_{i=-K}^K h_i
{\mathcal{E}(\gamma,\gamma_i) \over \mathcal{E}(\gamma_i)} ~ .
\label{thywmdcq}
\end{equation}
Equating expression (\ref{thywmdcq}) to $1$ for the $2K+1$ values (\ref{42})
yields the set of $2K+1$ linear equations:
\begin{equation}\label{47}
\text{E}\left[\hat{d}(\Theta,\Phi)|\gamma_j\right]= 1 \quad \Rightarrow \quad
\sum_{i=-K}^K \varepsilon_{i,j}\, h_i =1 ~ , \quad
\varepsilon_{i,j} = {\mathcal{E}(\gamma_j,\gamma_i) \over
\mathcal{E}(\gamma_i)}~ .
\end{equation}

\begin{quote}
\centerline{
\includegraphics[width=9cm]{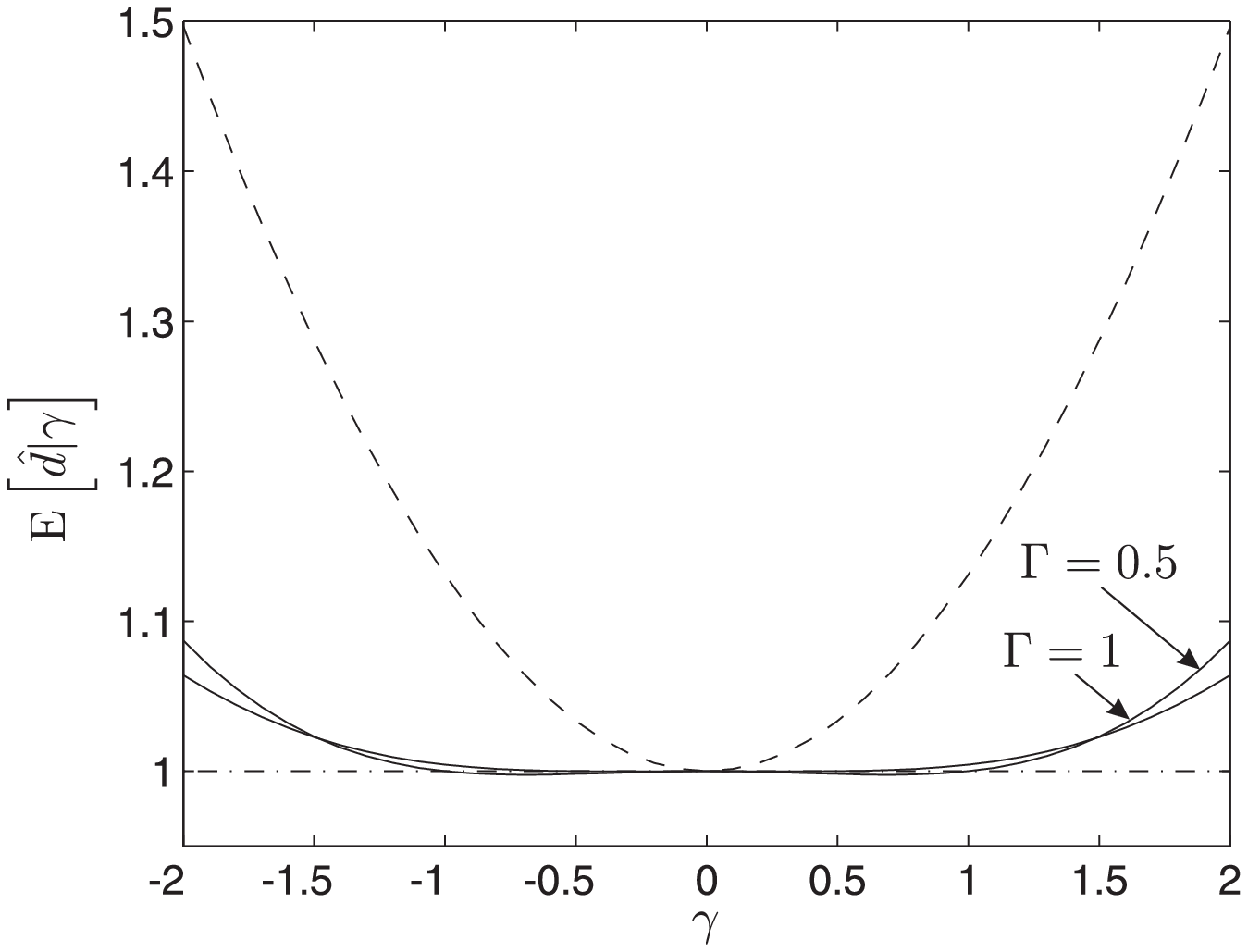}}
{\bf Fig.~8:} \small{Dependence as a function of $\gamma$ of the expected values of the R\&S (dash-dot line),
G\&K (dashed line) and of the quasi-unbiased first-order variance estimators
for the band widths $\Gamma=0.5; 1$ (solid lines).}
\end{quote}

\begin{quote}
\centerline{
\includegraphics[width=9cm]{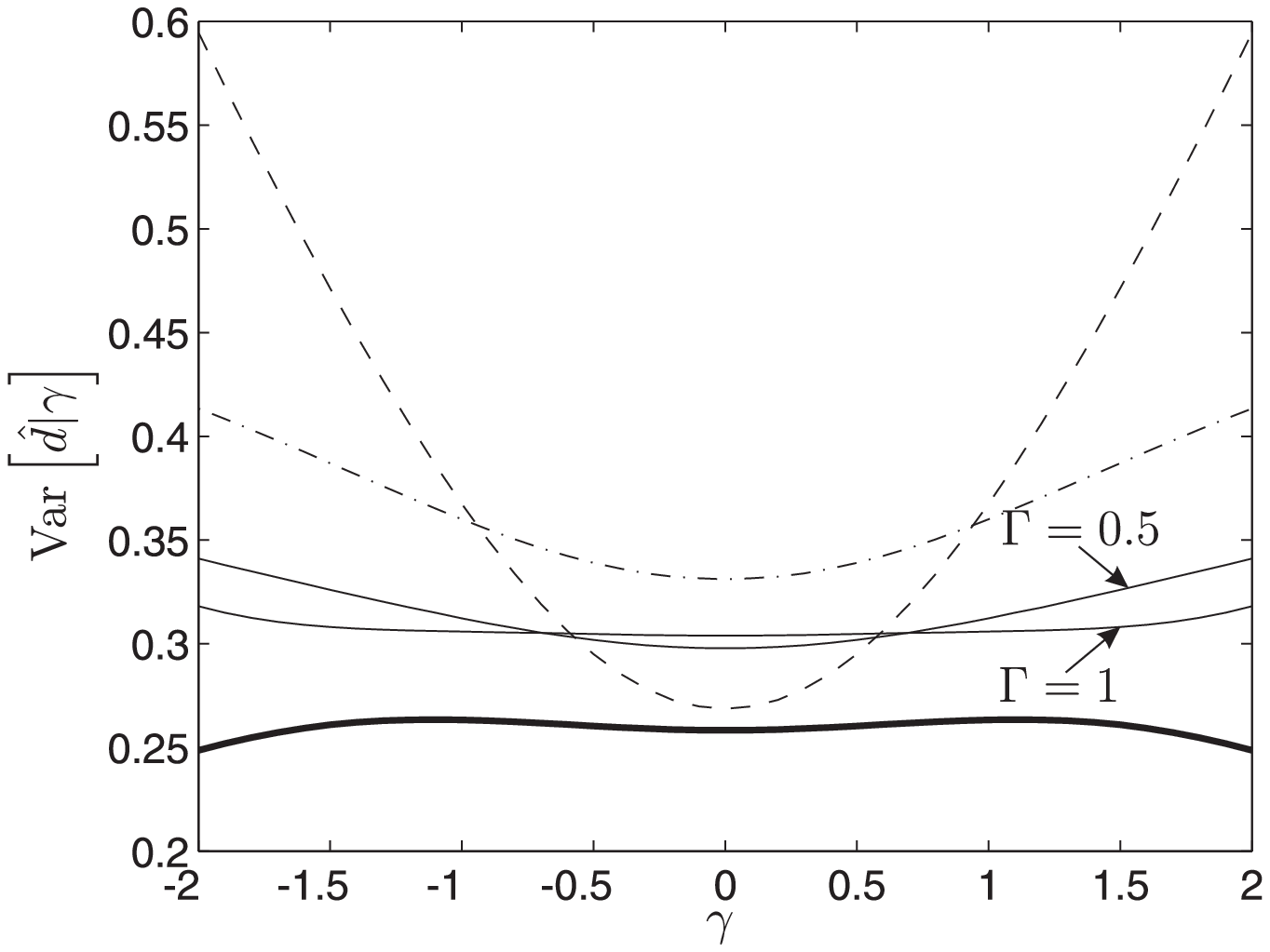}}
{\bf Fig.~9:} \small{Dependence as a function of $\gamma$ of the variances of the R\&S (dash-dot line),
G\&K (dashed line) and of the quasi-unbiased first-order variance estimators
for the band widths $\Gamma=0.5; 1$ (solid lines).}
\end{quote}

The statistical symmetry of the Wiener process \eqref{vtwiener} implies that the solution of equations \eqref{47} satisfies the following symmetry conditions:
$\varepsilon_{i,j} = \varepsilon_{-i,-j}$ $\Rightarrow$ $h_i = h_{-i}$.

Exploiting this symmetry for the first order case $K=1$ yields to two equations
for $h_1=h_{-1}$ and $h_0$:
$$
h_0 \varepsilon_{0,1} + h_1 (\varepsilon_{-1,1}+\varepsilon_{1,1}) = 1 ~ , \quad h_0 \varepsilon_{0,0} + 2 h_1 \varepsilon_{1,0} = 1 ~,
$$
whose solution reads
$$
h_0= {2\varepsilon_{1,0}- \varepsilon_{-1,1}- \varepsilon_{1,1} \over 2 \varepsilon_{0,1} \varepsilon_{1,0} -\varepsilon_{0,0}(\varepsilon_{-1,1} + \varepsilon_{1,1})}~ ,
\quad h_{\pm 1} = {\varepsilon_{0,0}- \varepsilon_{0,1} \over
\varepsilon_{0,0}(\varepsilon_{-1,1}+\varepsilon_{1,1}) -2
\varepsilon_{0,1} \varepsilon_{1,0}} ~ .
$$

Figure~8 shows the dependence as a function of $\gamma$ of the expected values of the first-order
quasi-unbiased canonical variance estimators for band widths $\Gamma= 0.5; 1$.
Figure~9 presents the dependence as a function of $\gamma$ of the
variances of these estimators. For comparison, the expected values and variances of the R\&S and G\&K estimators are also shown. We can state that the quasi-unbiased canonical variance estimators constructed here
provide the best of both world: (i) they exhibit a very weak bias up to rather large values of $\gamma$, thus competing
reasonably well with the R\&S  estimator; (ii) their variance is very weakly dependent on $\gamma$
and significantly smaller than that of the R\&S  estimator for all $\gamma$'s and than that of the G\&K   estimator, except
for a central zone around $\gamma=0$.

\section{Tests of theoretical results of variance and volatility estimators using synthetic time series of the Wiener process}

The present section implements the variance and volatility estimators discussed above
for synthetic time series of the Wiener process \eqref{xwindrift}. Because our results are mathematically exact,
these tests on synthetic time series offer the opportunity to study the impact of finite size and
discreteness effects, and give the opportunity to study additional properties of the estimators.
We will also determine the Maximum Likelihood estimator for the variance and the volatility
and will compare them to the other estimators.
The homogeneity of the estimators under study allow us to restrict $\sigma$ to the value $1$ and to
construct time series on the unit time interval ($T=1$), without losing generality.
With these parameter values, we have $\mu=\gamma$, and $X(t)$ are replaced by $v(\tau,\gamma)$ given by \eqref{vtwiener}.

\subsection{Test on numerical convergence of the discrete to the continuous Wiener process}

It is interesting to illustrate and test the theoretical results of previous sections by numerical simulation of the Wiener process $X(t)$ given by \eqref{xwindrift}.  Numerical simulations require replacing the continuous time stochastic process $v(\tau,\gamma)$ given by \eqref{vtwiener} by its discrete counterpart $v(n,\gamma)$ given by \eqref{vdiscret}, where $\{\epsilon_k\}$ are Gaussian.

The Gaussian discrete process $v(n,\gamma)$ represents rather accurately  the continuous time process $v(\tau,\gamma)$ only for sufficiently large $N$. On the other hand, the discrete process \eqref{vdiscret} obtained for not too large $N$ might describe the stochastic behavior of some financial markets more adequately than the continuous time process $v(\tau,\gamma)$. From a practitioner point of view,  $N$ could be interpreted as the typical number of transactions within the time interval of interest. From a theoretical point of view, $N$ should be chosen large enough to simulate the variables $\bar{H}$, $\bar{L}$ and $\bar{C}$ defined by \eqref{hlcbar}, which are known to be distributed according to the analytically derived pdf \eqref{qfh} with \eqref{condqhl}. To determine the appropriate value for $N$, we repeated $M=10^6$ simulations of the discrete process $v(n,\gamma)$ \eqref{vdiscret}, and calculated for each of these $M$ samples the corresponding G\&K and R\&S variance estimators at $\gamma=0$. Averaging over the $M$ realizations, we found the dependence of the expected value of the G\&K and R\&S variance estimators as a function of $N$, which is shown in figure~10. In particular, the statistical average value of the canonical R\&S estimator,  for $N=10^6$, is found to be $\text{E}\left[\hat{d}_\text{RS}\right|\gamma=0]=0.9987$, which is close enough to the theoretical one ($\text{E}\left[\hat{d}_\text{RS}\right|\gamma=0]=1$).

\begin{quote}
\centerline{
\includegraphics[width=9cm]{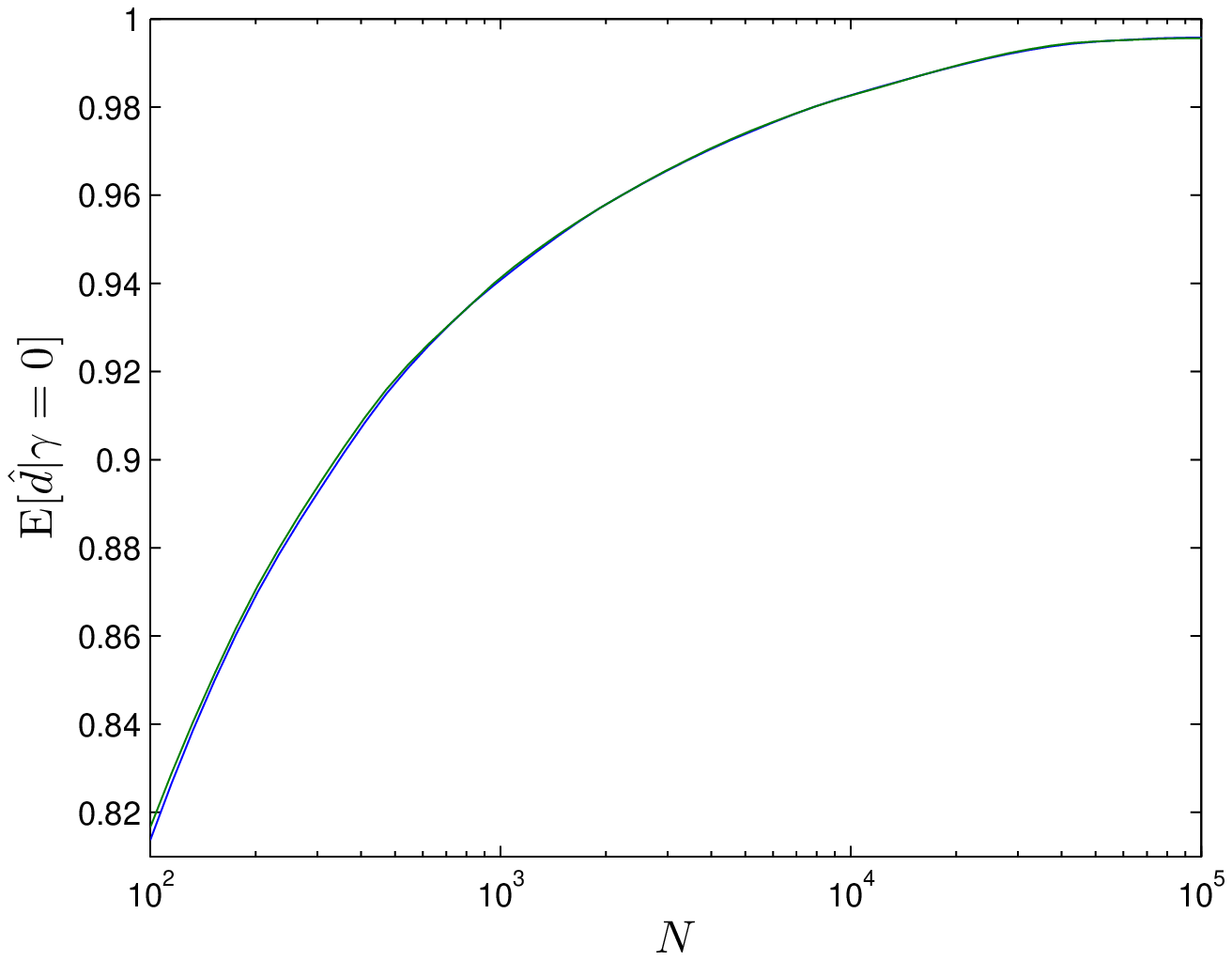}}
{\bf Fig.~10:} \small{Dependence as a function of $N$ of the statistical average value of the G\&K and  R\&S variance estimators for $\gamma=0$, where the statistical average is performed over $M=10^6$ realizations of the discrete time Wiener process $v(\tau,\gamma)$ given by \eqref{vtwiener}. Note that the two curves are almost undistinguishable, but not exactly the same.}
\end{quote}

\subsection{Variance estimators}

Figures~11 and~12 show the expected values and variances of the G\&K, R\&S and of the most efficient variance estimators, obtained theoretically and by numerical simulations with $M=10^5$ realizations of $v(n,\gamma)$, each of length $N=10^6$. One can observe an excellent agreement between the simulations and the theory.

\begin{quote}
\centerline{
\includegraphics[width=8.5cm]{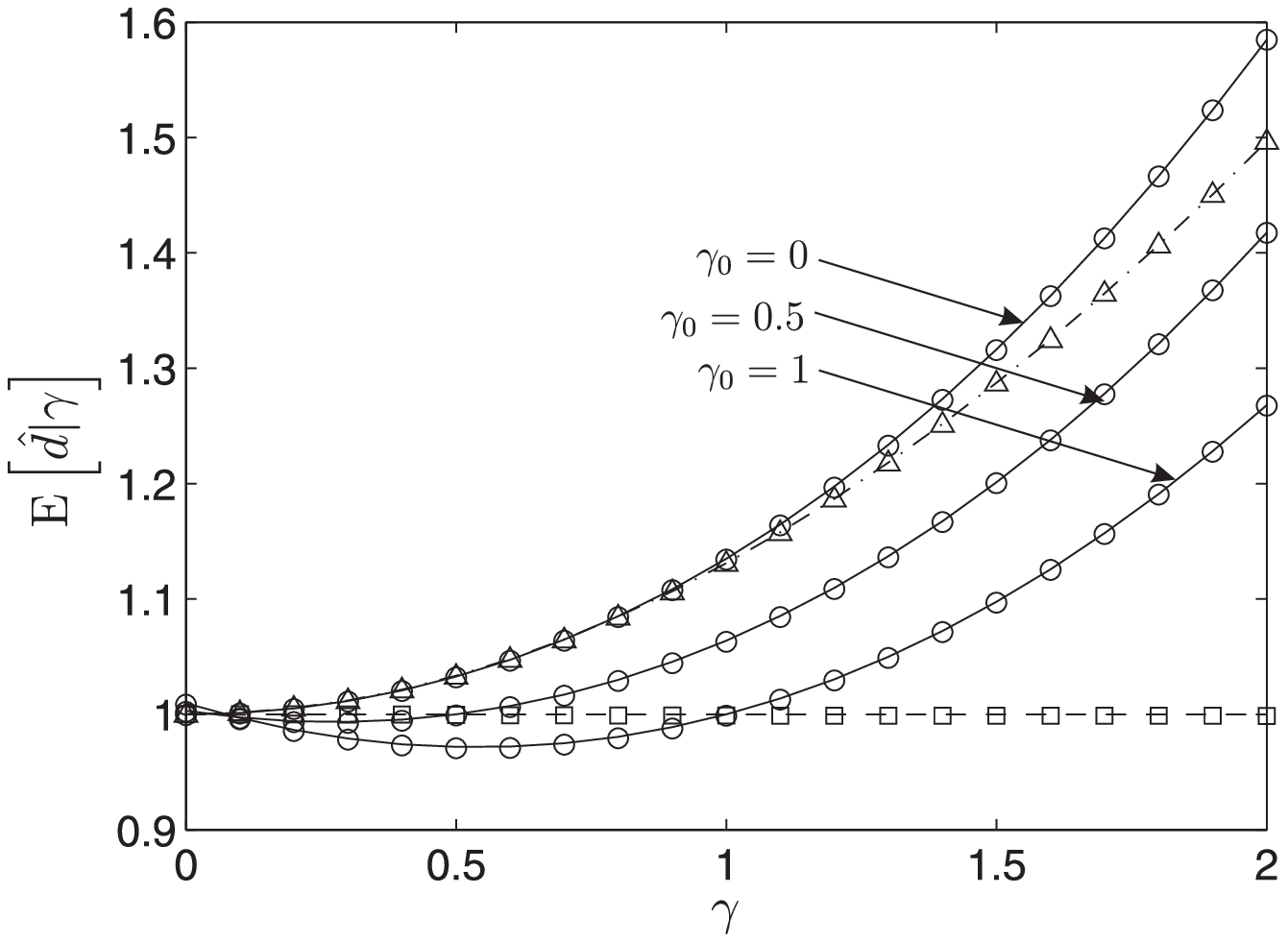}}
{\bf Fig.~11:} \small{Dependence of the expected values of the
R\&S (squares and dashed line), G\&K (triangles and dash-dot line)
and of the most efficient (at $\gamma_0=0;0.5;1$) (circles and solid lines) variance estimators as a function of $\gamma$. The markers show the values obtained by numerical simulations described in the text; the continuous lines correspond to the theoretical results presented in sections 3 and 4.}
\end{quote}

\begin{quote}
\centerline{
\includegraphics[width=8.5cm]{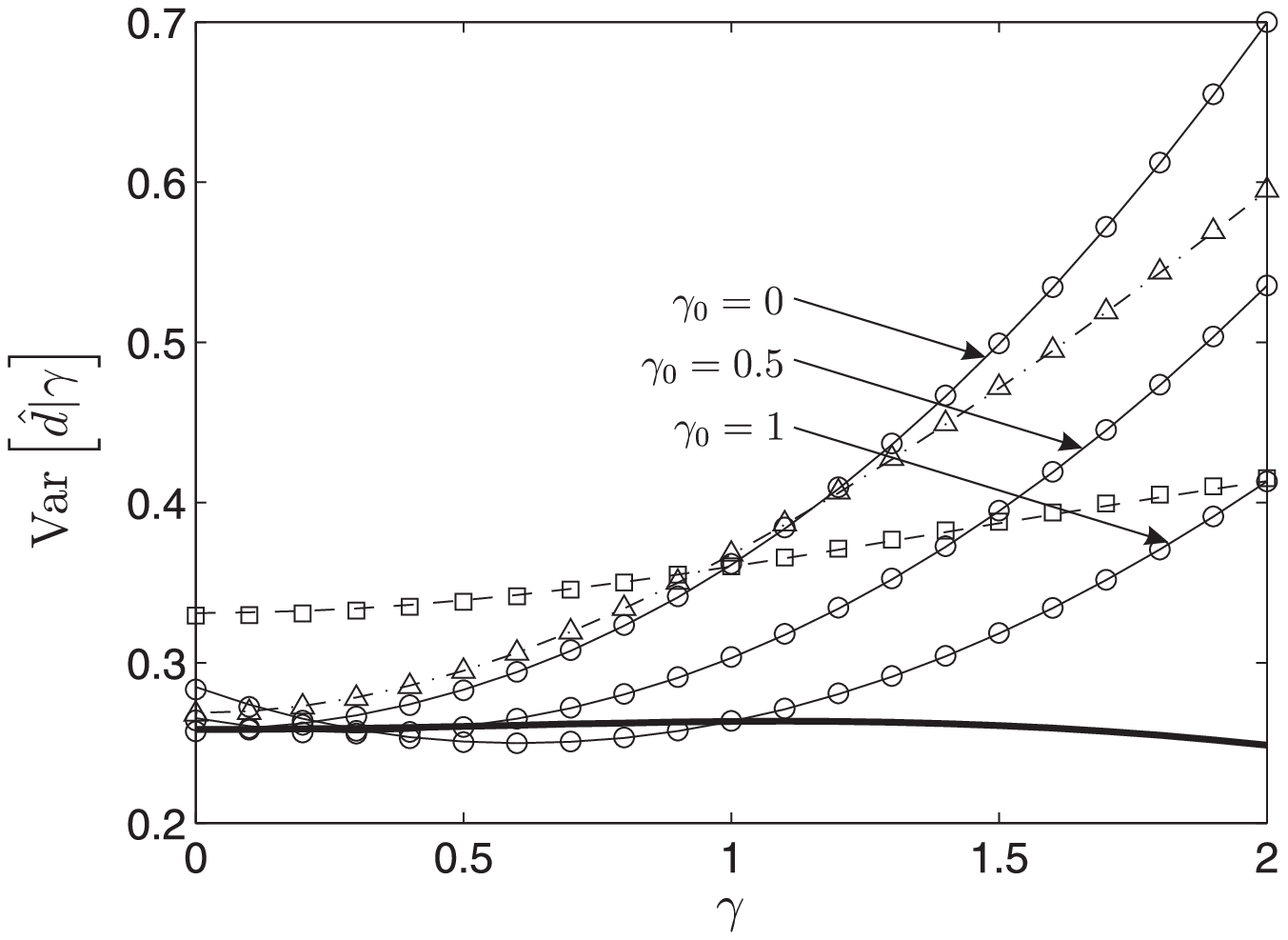}}
{\bf Fig.~12:} \small{Dependence of the variances of
the R\&S (squares and dashed line), G\&K (triangles and dash-dot line) and of the most efficient estimators (at $\gamma_0=0;0.5;1$) (circles and solid lines) of the variance estimators as a function
of $\gamma$. The markers show the values obtained by numerical simulations described in the text; the continuous lines correspond to the theoretical results presented in sections 3 and 4. }
\end{quote}

\subsection{Volatility estimators}

We now compare the efficiency and bias of the R\&S, G\&K and of the most efficient canonical volatility estimators.
Recall that, while the R\&S canonical variance estimator \eqref{Dtod} is unbiased for all $\gamma$'s,
the R\&S canonical volatility estimator \eqref{RSvolhom} is biased for all $\gamma$'s. The same holds true for the G\&K volatility estimator, which is biased even for $\gamma=0$. Figure~13 shows the dependence of the expected values of these estimators as a function of $\gamma$. In particular, the G\&K and R\&S volatility estimators have the following biases at $\gamma=0$:
$$
1- \text{E}[\hat{s}_\text{GK}|\gamma=0] = 0.0309~ , \qquad 1-\text{E}[\hat{s}_\text{RS}|\gamma=0] = 0.0386~ .
$$

\begin{quote}
\centerline{
\includegraphics[width=9cm]{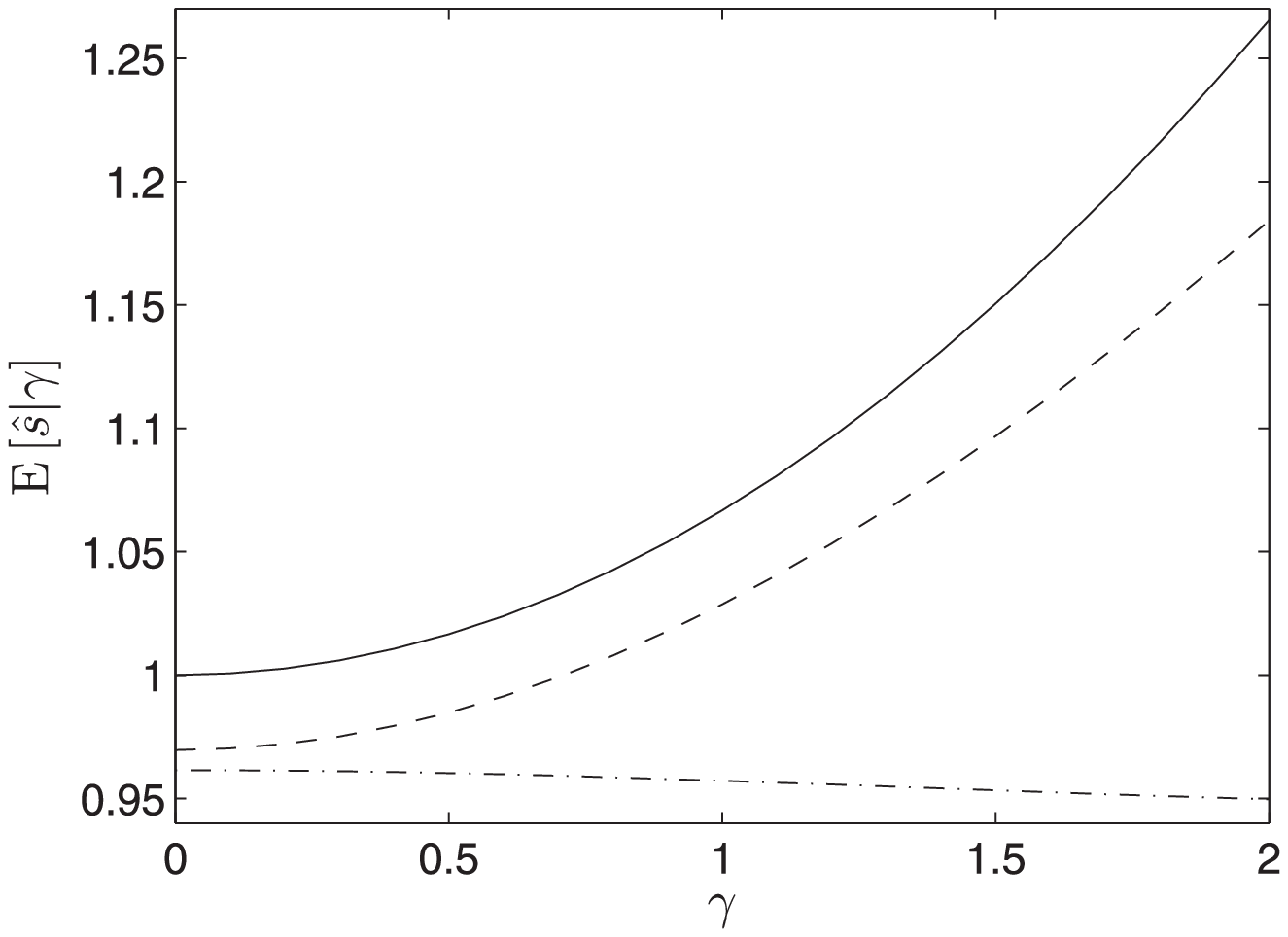}}
{\bf Fig.~13:} \small{Dependence as a function of $\gamma$ of the expected values of
R\&S (dash-dot), G\&K (dashed) and of the most efficient at $\gamma_0=0$ volatility estimator (solid line).}
\end{quote}

In order to provide an appropriate comparison between the efficiency of the R\&S, G\&K and of the most
efficient volatility estimators, we normalize them by their values reached at $\gamma=0$:
\begin{equation}
\hat{s}_\text{norm}(\Theta,\Phi) = \bar{R}
\psi(\Theta,\Phi)\big/\text{E}[\hat{s}|\gamma=0]~ .
\label{thywgf}
\end{equation}

\begin{quote}
\centerline{
\includegraphics[width=10cm]{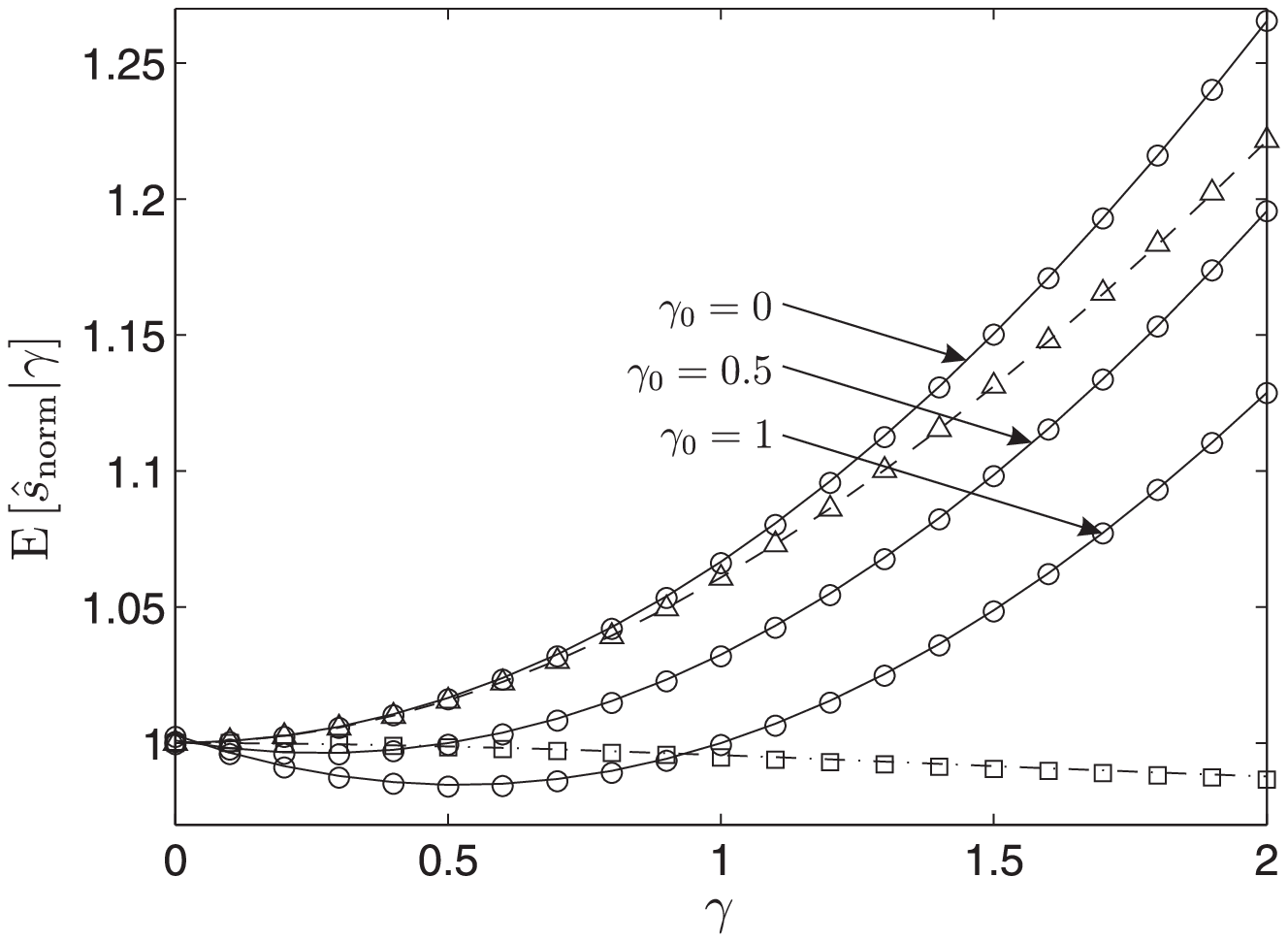}}
{\bf Fig.~14:} \small{Expected values of the normalized (according to (\ref{thywgf})) R\&S, G\&K and of the most efficient unbiased homogeneous canonical volatility estimators at $\gamma_0=0;0.5;1$.}
\end{quote}

\begin{quote}
\centerline{
\includegraphics[width=10cm]{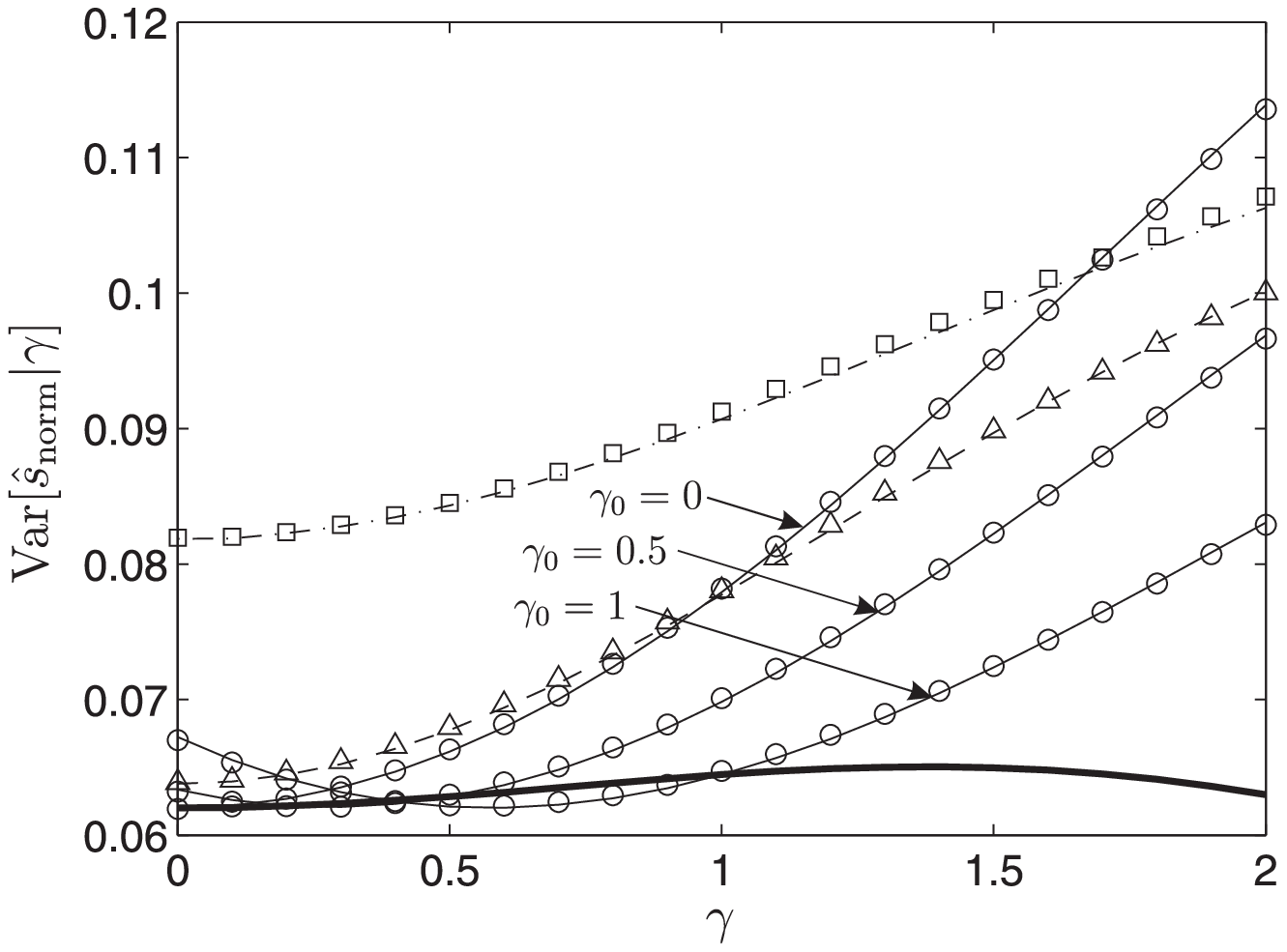}}
{\bf Fig.~15:} \small{Variances of the normalized (according to (\ref{thywgf})) R\&S, G\&K and of the most efficient unbiased homogeneous canonical volatility estimators at $\gamma_0=0;0.5;1$. The heavy solid line corresponds to the lowest bound variance $W(\gamma)$ given by \eqref{wgamleastvoldef}.}
\end{quote}

Figures~14 and 15 show the expected values and variances of the normalized (according to (\ref{thywgf})) R\&S, G\&K and of the most efficient unbiased homogeneous canonical volatility estimators at $\gamma_0=0;0.5;1$. In particular, the variances at $\gamma=0$ of the normalized G\&K, R\&S and of the most efficient (at $\gamma_0=0$) volatility estimators are equal to
\begin{equation}\label{volmersgkdatas}
\begin{array}{c}
\text{Var}[\hat{s}_\text{GK}|\gamma=0] = 0.06379~ , \quad \text{Var}[\hat{s}_\text{RS}|\gamma=0] = 0.08186~ , \\
\text{Var}[\hat{s}(\gamma_0=0)|\gamma=0] = 0.06201~ .
\end{array}
\end{equation}
The theoretical results shown in figures~14 and 15 are also compared with the numerical calculations performed
using $M=10^5$ different realizations of the discrete
Wiener process \eqref{vdiscret} with length $N=10^6$..

\subsection{Maximum likelihood estimators}

The Appendix derives the exact expression for the joint distribution of the $H, L, C$ of a Wiener process.
Being a function of the volatility $\sigma$, this joint distribution allows us to obtain the maximum likelihood (ML) estimator
of $\sigma$, as we now describe. It turns out that the MLE is less efficient than the most efficient homogenous
estimators described above.

Let us start from $\bar{\mathcal{Q}}(h,l,c;\gamma)$ given by \eqref{qfh} in the Appendix, which is the pdf of the high, low and close values $(\bar{H},\bar{L}, \bar{C})$ defined by \eqref{hlcbar} of the Wiener process $v(\tau,\gamma)$ \eqref{vtwiener} with unit volatility. Knowing  $\bar{\mathcal{Q}}(h,l,c;\gamma)$, one can recover the pdf  $\mathcal{Q}(\eta,\lambda,\xi;\mu,\sigma)$ of the high, low and close values $(H,L,C)$ defined by \eqref{hlcsup} of the original Wiener process $X(t)$ \eqref{xwindrift} for $t\in(0,T)$, by using the relation
\begin{equation}\label{qorighlc}
\begin{array}{c} \displaystyle
\mathcal{Q}(\eta,\lambda,\xi;\mu,\sigma) ={1 \over \sigma^3 T \sqrt{T}} \bar{\mathcal{Q}}\left({\eta \over \sigma \sqrt{T}},{\lambda \over \sigma \sqrt{T}},{\xi \over \sigma \sqrt{T}}; {\mu \over \sigma} \sqrt{T}\right) =
\\[5mm] \displaystyle
{1 \over \sigma^3 T \sqrt{2\pi T}} \exp\left(-{(\xi-\mu T)^2 \over 2 \sigma^2 T}\right) \mathcal{R}\left({\eta \over \sigma \sqrt{T}},{\lambda \over \sigma \sqrt{T}}\bigg|{\xi \over \sigma \sqrt{T}}\right)~ .
\end{array}
\end{equation}
This expression for the pdf of $(H,L,C)$ allows us to construct the maximum likelihood OHLC estimators $\hat{\mu}_\text{ML}$ and $\hat{\sigma}_\text{ML}$ of the drift and the volatility of the Wiener process $X(t)$ defined by \eqref{xwindrift}. The MLE are obtained by replacing the arguments $(\eta,\lambda,\xi)$ in \eqref{qorighlc} by the realized samples $(H,L,C)$, and by
searching for the values $\hat{\mu}_\text{ML}$ and $\hat{\sigma}_\text{ML}$ that maximize the likelihood function
$$
\begin{array}{c}
\mathcal{L}(H,L,C;\mu,\sigma) = \ln \mathcal{Q}(H,L,C;\hat{\mu}, \hat{\sigma}) =
\\[4mm] \displaystyle
-{(C-\hat{\mu}_\text{ML} T)^2 \over 2 \hat{\sigma}_\text{ML}^2 T} +
\ln \mathcal{R}\left({H \over \hat{\sigma}_\text{ML} \sqrt{T}},{L \over \hat{\sigma}_\text{ML} \sqrt{T}}\bigg|{C \over \hat{\sigma}_\text{ML} \sqrt{T}}\right) - 3 \ln \hat{\sigma}_\text{ML}~ .
\end{array}
$$
We obtain the ML drift estimator,
\begin{equation}
\hat{\mu}_\text{ML} = {C \over T}~ .
\label{yhyuj4uksl}
\end{equation}
We recall that this drift estimator (\ref{yhyuj4uksl})
has the minimal possible variance among all estimators, since it
realizes the lower bound given by the Cramer-Rao inequality.

The ML volatility estimator $\hat{\sigma}_\text{ML}$ maximizes the function
\begin{equation}\label{mathnvolest}
\ln \mathcal{R}\left({H \over \hat{\sigma}_\text{ML} \sqrt{T}},{L \over \hat{\sigma}_\text{ML} \sqrt{T}}\bigg|{C \over \hat{\sigma}_\text{ML} \sqrt{T}}\right) - 3 \ln \hat{\sigma}_\text{ML}~ .
\end{equation}
The following theorem then derives.
\begin{theorem}\label{thmlhom}
The ML volatility estimator $\hat{\sigma}_\text{ML}$ is homogeneous, i.e., analogously to \eqref{varcanestthph}, it
can written in the form
$$
\hat{\sigma}_\text{ML} = \sigma \hat{s}_\text{ML}(\bar{H}, \bar{L}, \bar{C})~,
$$
where $\hat{s}_\text{ML}(h, l, c)$ is a first order homogeneous function.
\end{theorem}

\emph{\textbf{Proof.}}
Replacing $\hat{\sigma}_\text{ML}$ by
$\hat{\sigma}_\text{ML} = \sigma \hat{s}_\text{ML}$ in expression \eqref{mathnvolest}, using the equalities
$$
\bar{H} = {H \over \sigma \sqrt{T}}~ , \quad \bar{L}= {L \over \sigma \sqrt{T}}~ , \quad \bar{C} = {C \over \sigma \sqrt{T}}~ ,
$$
and omitting the nonessential constant $3\ln\sigma$, we obtain that $\hat{s}_\text{ML}$ should maximize the function
\begin{equation}\label{nfuncmlevolest}
\mathcal{N}(\bar{H}, \bar{L}, \bar{C}, \hat{s}_\text{ML}) =
\ln \mathcal{R}\left({\bar{H} \over \hat{s}_\text{ML}} ,{\bar{L} \over \hat{s}_\text{ML} }\bigg|{\bar{C} \over \hat{s}_\text{ML}}\right)- 3 \ln \hat{s}_\text{ML}~ .
\end{equation}
Here, $\mathcal{R}(h,l|c)$ is a deterministic function given by \eqref{condqhl}. Accordingly, the value $\hat{s}_\text{ML}$, which maximizes the function $\mathcal{N}(\bar{H}, \bar{L}, \bar{C}, \hat{s}_\text{ML})$, is a deterministic function $\hat{s}_\text{ML} = \hat{s}_\text{ML}(\bar{H}, \bar{L}, \bar{C})$ of the variables $(\bar{H}, \bar{L}, \bar{C})$. Its homogeneity is obvious.  $\square$

\begin{rem}
From general properties of maximum likelihood estimators, the ML variance estimator is also homogeneous and it is equal to the square of the volatility estimator:
\begin{equation}\label{mlevarest}
\hat{D} = \sigma^2 \hat{d}_\text{ML}(\bar{H}, \bar{L}, \bar{C})~ , \qquad
\hat{d}_\text{ML}(\bar{H}, \bar{L}, \bar{C}) = \hat{s}^2_\text{ML}(\bar{H}, \bar{L}, \bar{C})~ .
\end{equation}
\end{rem}

In general, ML estimators are biased. It is therefore convenient to normalize it by its value as
some given $\gamma=\gamma_0$ to obtain
$$
\hat{s}_\text{norm} = {\hat{s}_\text{ML}(\bar{H}, \bar{L}, \bar{C}) \over \text{E}[\hat{s}_\text{ML}(\bar{H}, \bar{L}, \bar{C})|\gamma_0]}~.
$$
Since ML estimators are homogeneous, they may not be more efficient than the most efficient estimators at the same $\gamma_0$ value. In practice, unbiased ML estimators are significantly less efficient than the most efficient one.
Let illustrate this fact using the normalized ML volatility estimator at $\gamma_0=0$.
For this case, the numerical calculation with $(N=10^6$, $M=10^6)$ of the expected value and variance, at $\gamma=0$, of the canonical ML estimator yields
\begin{equation}
\text{E}[\hat{s}_\text{ML}|0] \approx 0.9202~ , \quad \text{Var}[\hat{s}_\text{ML}|0] \approx 0.0712 \quad \Rightarrow \quad \text{Var}[\hat{s}_\text{norm}|0] \approx 0.0840~ .
\end{equation}
Comparing these values with those reported in \eqref{volmersgkdatas}, one can see that the efficiency of the ML volatility estimator is significantly worse than for the most efficient one, and even worse than that of the R\&S volatility estimator. The corresponding values for the ML canonical variance estimator are
\begin{equation}
\text{E}[\hat{d}_\text{ML}|0] \approx 0.9179~ , \quad \text{Var}[\hat{d}_\text{ML}|0] \approx 0.2756 \quad \Rightarrow \quad \text{Var}[\hat{d}_\text{norm}|0] \approx 0.3271~ .
\end{equation}
While smaller than the variance of the R\&S canonical variance estimator ($\text{Var}_\text{RS}[\hat{d}|0]\approx 0.331$), the variance $\text{Var}[\hat{d}_\text{norm}|0]$  is 27\% larger than the variance of the most efficient one ($V(0)\approx 0.258$).

\section{Conclusions}

We have laid the first stones for a comprehensive theory of homogeneous volatility (and variance) estimators of
arbitrary stochastic processes. Our focus has been to exploit the universally quoted OHLC (open-high-low-close)
prices, which can span time intervals extending from seconds to years, in order to develop new efficient estimators.
Our theory opens many possibilities to design new efficient estimators, such as the ``quasi-unbiased estimators'',
that address any type of desirable constraints. The main tool of our theory is the parsimonious encoding
of all the information contained in the OHLC in the form of general ``diagrams'' associated with the
joint distributions of the high minus open, low minus open and close minus open values. The diagrams
can be tailored to yield the most efficient estimators associated to any statistical properties of the underlying log-price stochastic process.

Our theory opens several interesting developments. First, the accurate determination of the key functions $g_n(\theta,\phi;\gamma)$, defining the above diagrams, gives the tools to develop efficient estimators of the variance and volatility (as well as any other quantities of interest)  for arbitrary non-Gaussian log-price processes, including the presence of
micro-structure as in tick-by-tick price series. Our methods should lead to the development of fast and effective algorithms for low- and high-frequency OHLC variance and volatility estimators, that can be applied in practice to any kind of financial markets.

\clearpage

\appendix
\setcounter{section}{0}
\setcounter{equation}{0}
\renewcommand{\theequation}{\thesection.{\arabic{equation}}}
\renewcommand{\thesection}{\Alph{section}}

\section{Extremes of Wiener processes}

In the main text, we lay out
the basic stones for a comprehensive theory of homogenous
OHLC volatility and variance estimators, which are most efficient for any specific value
of the normalized drift parameter $\gamma$ of the underlying price stochastic process.
This theory uses the OHLC (open-high-low-close) prices in the given time interval or scale
of interest.

All expressions depend on a fundamental quantity, which is the
joint probability density function (pdf) $\bar{\mathcal{Q}}(h,l,c;\gamma)$ defined by \eqref{pdfhlcdef} of the high, low and close values given by \eqref{hlcbar} of the auxiliary stochastic process $B(t,\gamma)$ \eqref{xtoy}.

In general, it is only possible to construct the sought pdf $\bar{\mathcal{Q}}(h,l,c;\gamma)$ by numerical simulations generating a huge number of realizations of the underlying stochastic process $B(t,\gamma)$. For certain
stochastic process $X(t)$ \eqref{genmod}, the pdf $\bar{\mathcal{Q}}(h,l,c;\gamma)$ can be calculated analytically.
In this Appendix, we obtain the explicit analytical expression for $\bar{\mathcal{Q}}(h,l,c;\gamma)$ in the case of the Wiener process, $B(t,\gamma)\equiv v(\tau,\gamma)$ given by expression \eqref{vtwiener}.

As shown below, the sought pdf $\bar{\mathcal{Q}}(h,l,c;\gamma)$ will be derived from the solution of the diffusion equation
\begin{equation}\label{difeqv}
{\partial f(c;\tau,\gamma) \over \partial \tau} + \gamma {\partial f(c;\tau,\gamma) \over \partial c} = {1 \over 2} {\partial^2 f(c;\tau,\gamma) \over \partial c^2} ~ ,
\end{equation}
where the reduced time $\tau$ and parameter $\gamma$ are defined in \eqref{gamdef}.
The well-known solution of the diffusion equation \eqref{difeqv}, satisfying the initial condition
\begin{equation}\label{inconddifeq}
f(c;\tau = 0,\gamma) = \delta(c)~ ,
\end{equation}
is
\begin{equation}\label{fdifeqsol}
f(c;\tau,\gamma) = g(c-\gamma \tau, \tau)~ , \qquad g(x,\tau) =
{ 1 \over \sqrt{2\pi \tau}} \exp\left( - {x^2 \over 2 \tau} \right)~ .
\end{equation}

\subsection{Distribution of the maximal value}

The full derivation of the pdf $\bar{\mathcal{Q}}(h,l,c;\gamma)$ for the Wiener process $v(\tau,\gamma)$ \eqref{vtwiener} involves rather extensive calculations. In order to present the intuition behind these calculations, it is useful to
consider the reduced problem of determining the joint pdf of the high (maximum) and close values of the Wiener process $v(\tau';\gamma)$ within a given time interval $\tau'\in(0,\tau)$. This reduced problem is tightly connected with the so-called ``absorption'' of the process $v(\tau;\gamma)$ at the given level $h$. The existence of absorption amounts to supplement the diffusion equation \eqref{difeqv} by the absorption condition
\begin{equation}\label{absboundcond}
f(c=h;\tau,\gamma) = 0~ , \qquad h>0~ .
\end{equation}

We denote the solution of the initial-boundary value problem \eqref{difeqv}, \eqref{inconddifeq}, \eqref{absboundcond} by $f(c,h;\tau,\gamma)$. This function is the pdf of the values, at time $\tau$, of the realizations of the stochastic process $v(\tau';\gamma)$, that has not reached the level $h$ for all times $\tau'\in(0,\tau)$, i.e.,
\begin{equation}\label{dabsorbdist}
f(c,h;\tau,\gamma) dx = \Pr\{v(\tau;\gamma)\in(x,x+dx) \cap \bar{H} < h \}~, \quad x<h~ , h>0~ ,
\end{equation}
where
$$
\bar{H} = \sup_{\tau'\in(0,\tau)} v(\tau',\gamma)~ .
$$
Correspondingly, expression \eqref{dabsorbdist} implies that the joint pdf of the random variables $\bar{C}=v(\tau,\gamma)$ and maximum $\bar{H}$ is equal to
\begin{equation}
\bar{\mathcal{Q}}(h,c;\gamma,\tau) = {\partial f(c,h;\tau,\gamma) \over \partial h}~ , \qquad h> 0~ , \quad c<h~ .
\label{ntynjymjyrj}
\end{equation}
Then, the joint pdf of the high and close values of the stochastic process $v(\tau',\gamma)$ within the interval $\tau'\in(0,1)$
is obtained by taking $\tau=1$ in expression (\ref{ntynjymjyrj}), which reads
\begin{equation}\label{jointhcdef}
\bar{\mathcal{Q}}(h,c;\gamma) = {\partial f(c,h;\tau=1,\gamma) \over \partial h}~ , \qquad h> 0~ , \quad c<h~ .
\end{equation}
The joint pdf of the high and close values of Brownian motions was derived by Paul L\'evy (1948).

The solution of the initial-boundary value problem \eqref{difeqv}, \eqref{inconddifeq}, \eqref{absboundcond} can
be obtained by the \emph{reflection method} as follows. The reflection method consists in replacing the initial-boundary value problem by the following auxiliary initial-value problem
\begin{equation}\label{incondrefone}
\begin{array}{c}\displaystyle
{\partial f(c,h;\tau,\gamma) \over \partial \tau} + \gamma {\partial f(c,h;\tau,\gamma) \over \partial c} = {1 \over 2} {\partial^2 f(c,h;\tau,\gamma) \over \partial c^2} ~ ,
\\[4mm] \displaystyle
f(c,h;\tau=0,\gamma) = \delta(c) - A \delta(c-2 h) ~,
\end{array}
\end{equation}
where the constant $A$ has to be chosen such that the solution of the initial-value problem \eqref{incondrefone} satisfies the absorbtion boundary condition \eqref{absboundcond}.

The solution of the initial value problem \eqref{incondrefone} is nothing but
$$
f(c,h;\tau,\gamma) = g(c-\gamma \tau,\tau) - A g(c-2h -\gamma \tau,\tau)~ ,
$$
where $g(x,\tau)$ is given in \eqref{fdifeqsol}. Substituting this expression into the boundary condition \eqref{absboundcond}
yields $A = e^{2 h \gamma}$. Thus, the solution of the initial-boundary value problem is
\begin{equation}\label{solinvponebev}
f(c,h;\tau,\gamma) = g(c-\gamma \tau,\tau) - e^{2 h \gamma} g(c-2h -\gamma \tau,\tau)~ .
\end{equation}
Substituting it into expression \eqref{jointhcdef} yields the joint pdf of the high and close variables,
\begin{equation}\label{hcjointd}
\bar{\mathcal{Q}}(h,c;\gamma) = f(c;\gamma) \mathcal{R}(h|c)~ , \qquad c<h~ , \qquad h>0~ ,
\end{equation}
where
\begin{equation}\label{fpdfctone}
f(c;\gamma) = {1 \over \sqrt{2 \pi}} \exp\left(-{(c-\gamma)^2 \over 2} \right)
\end{equation}
is the pdf of the close value $c=v(1,\gamma)$, while
$$
\mathcal{R}(h|c) = 2 (2h-c) e^{2 h(c-h)}~ , \qquad h\geqslant \max\{0,c\} ,
$$
is the pdf of the high value $\bar{H}$, under the condition that the close value is equal to $c$.

\subsection{Wiener process between two absorbing boundaries}

The joint pdf $Q(h,l,c;\gamma)$ defined by \eqref{pdfhlcdef}  of the
high, low and close values of the Wiener process can be expressed similarly to relation
\eqref{jointhcdef} via the solution of the diffusion equation \eqref{difeqv} in the presence
of two absorbing boundaries. We thus the new initial-boundary problem
\begin{equation}\label{incondreftwo}
\begin{array}{c}\displaystyle
{\partial f(c,h,l;\tau,\gamma) \over \partial \tau} + \gamma {\partial f(c,h,l;\tau,\gamma) \over \partial c} = {1 \over 2} {\partial^2 f(c,h,l;\tau,\gamma) \over \partial c^2} ~ ,
\\[2mm] \displaystyle
f(c,h,l;\tau=0,\gamma) = \delta(c) ~,
\\[1mm] \displaystyle
f(c=h+u \tau,h,l;\tau,\gamma)= 0~ , \qquad f(c=l+ v \tau,h,l;\tau,\gamma)= 0~ .
\end{array}
\end{equation}
Using the reflection method and a derivation similar to that leading to
expression \eqref{solinvponebev}, we obtain
\begin{equation}\label{ftwoboundsol}
\begin{array}{c} \displaystyle
f(c,h,l;\tau,\gamma) =\sum_{m=-\infty}^\infty e^{2 (v-u) (m(h-l)+l)} \times
\\[2mm] \displaystyle
\big[ e^{2(v-\gamma)(h-l)m} g(c-\gamma \tau + 2(h-l) m,\tau) -
\\[2mm] \displaystyle
e^{2(\gamma-v)((h-l)m+l)} g(c-\gamma\tau -2 l - 2(h-l)m,\tau)\big]~ .
\end{array}
\end{equation}
Figure~A1 plots the function
\begin{equation}\label{fandfund}
f(c,h,l;\tau,\gamma) + 0.05 \cdot \tau~
\end{equation}
as a function of the close value $c$.
The $ 0.05 \cdot \tau$ is added in order to show clearly that $f(c,h,l;\tau,\gamma)$ indeed satisfies the moving absorption conditions \eqref{incondreftwo}.

We need the particular case corresponding to static boundaries ($u=v=0$) to transform
the general solution \eqref{ftwoboundsol} into
\begin{equation}\label{ftwoboundsolstatic}
\begin{array}{c} \displaystyle
f(c,h,l;\tau,\gamma) =\sum_{m=-\infty}^\infty
\big[ e^{2\gamma(l-h)m} g(c-\gamma \tau + 2(h-l) m,\tau) -
\\[4mm] \displaystyle
e^{2\gamma((h-l)m+l)} g(c-\gamma\tau -2 l - 2(h-l)m,\tau)\big]~ , \\[1mm]
l<c<h~ , \quad h>0~ , \quad l<0~ .
\end{array}
\end{equation}

\begin{quote}
\centerline{
\includegraphics[width=9cm]{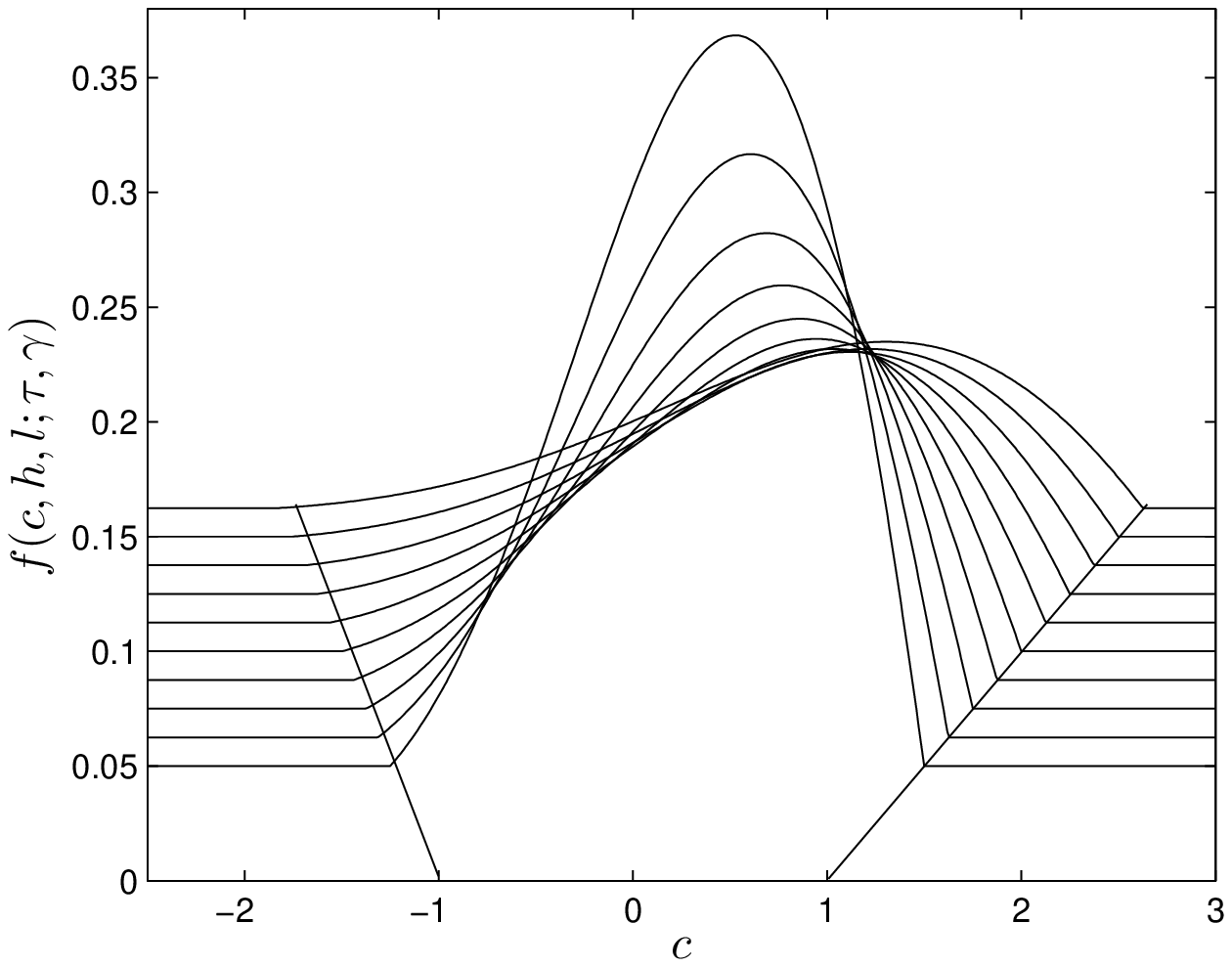}}
{\bf Fig.~A1:} \small{Plots of the function \eqref{fandfund} as a function of the close value $c$ for $h=1$, $l=-1$, $u=0.5$, $v=-0.25$, $\gamma=0.8$ and for $\tau=1+0.25\cdot k$, where $k=0,1,\dots,9$.}
\end{quote}

\subsection{Distribution of high, low, close values}

The joint pdf $\bar{\mathcal{Q}}(h,l,c;\gamma)$ corresponding to the diffusion process $v(\tau',\gamma)$ within the time interval $\tau'\in(0,1)$ is obtained via the pdf $f(c,h,l;\tau,\gamma)$ given by \eqref{ftwoboundsolstatic} by
the following relation, which is analogous to \eqref{jointhcdef}:
$$
\bar{\mathcal{Q}}(h,l,c;\gamma) = - {\partial f(c,h,l;\tau=1,\gamma) \over \partial h \partial l}~ .
$$
Analogously to expression \eqref{hcjointd}, we obtain
\begin{equation}\label{qfh}
\bar{\mathcal{Q}}(h,l,c;\gamma) = f(c;\gamma) \mathcal{R}(h,l|c) ~ , \quad h> 0~ , \quad l<0~ , \quad l<c<h~ ,
\end{equation}
where $f(c;\gamma)$ is given by \eqref{fpdfctone}, while $\mathcal{R}(h,l|c)$ is the joint pdf of the high and low values under the condition that the close value is equal to $c$:
\begin{equation}\label{condqhl}
\begin{array}{c}\displaystyle
\mathcal{R}(h,l|c) =
\\[2mm] \displaystyle
4 \sum_{m=-\infty}^\infty m
\big[ m \mathcal{D}(m (h-l),c) + (1-m) \mathcal{D}(m(h-l)+l,c) \big]~,
\\[2mm] \displaystyle
\mathcal{D}(h,c) = [(c-2 h)^2-1] e^{2 h (c-h)}~.
\end{array}
\end{equation}
Figure~A2 shows the contour lines of the conditional pdf $\mathcal{R}(h,l|c)$ for $c=0$
in the plane $(h, l)$. Skorohod (1964) reported the joint distribution of the high-low-close
for random walks with zero drift ($\gamma=0$).

\begin{quote}
\centerline{
\includegraphics[width=9cm]{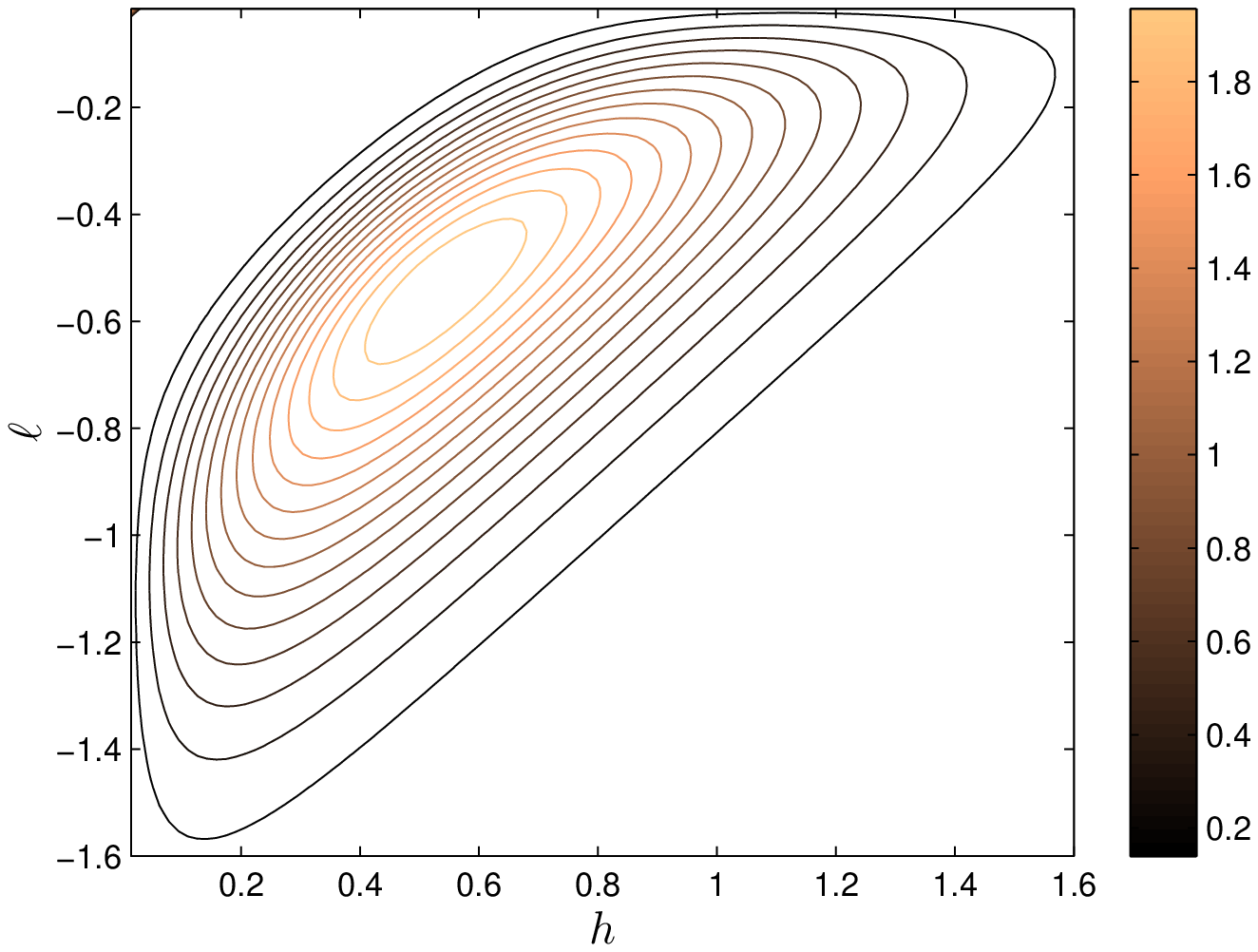}}
{\bf Fig.~A2:} \small{Contour lines of the conditional pdf $\mathcal{R}(h,l|c)$ given by \eqref{condqhl} for $c=0$
in the plane $(h, l)$.}
\end{quote}

\subsection{Function $g_n$ defined in expression \eqref{gnintrdef}}

As seen from expressions \eqref{mevardiagrest} and \eqref{mevardiagrestv}, the diagrams (see definition \ref{yjujifkd}) of the most efficient estimators are expressed via the function $g_n(\theta,\phi;\gamma)$ defined by the equation \eqref{gnintrdef}.
The above calculations valid for the Wiener process show that it is equal to
$$
\begin{array}{c}\displaystyle
g_n(\theta,\phi;\gamma) = {4 \over \sqrt{2\pi}} e^{-\gamma^2/ 2} ~\times
\\[3mm] \displaystyle
\sum_{m=-\infty}^\infty m
\big[ m I_n(m (\tilde{h}-\tilde{l}),\tilde{c};\gamma) + (1-m) I_n(m(\tilde{h}-\tilde{l})+\tilde{l},\tilde{c};\gamma) \big]~,
\end{array}
$$
where
$$
I_n(h,c,\gamma) = \int_0^\infty \rho^{2+n} \exp\left(\gamma c\rho-{c^2 \over 2}\rho^2\right) \mathcal{D}(h\rho,c\rho) d\rho~
$$
and
$$
\tilde{h}= \cos\theta \cos\phi~ , \qquad \tilde{l} = \cos\theta\sin\phi~ , \qquad \tilde{c} = \sin\theta~ .
$$
In particular,
$$
I_n(h,c,\gamma=0) = {F(n) \over |2h-c|^{3+n}} ~ , \qquad F(n) = 2^{1+n \over 2} (2+n) \Gamma\left({3+n \over 2}\right)~ .
$$

\end{document}